\documentclass{article}
\usepackage{ifpdf}
\ifpdf\fi
\usepackage{fullpage}
\usepackage{url}
\usepackage{times}
\usepackage{subfigure}
\usepackage{algorithm}
\usepackage[noend]{algorithmic}
\usepackage{url}
\usepackage{amssymb}
\usepackage{amsmath}
\usepackage{amsfonts}
\usepackage{latexsym}
\usepackage{graphicx}
\usepackage{mathptmx}
\usepackage{epstopdf}

\newcommand{\para}[1]{\medskip \noindent {\bf #1}}

\newtheorem{theorem}{Theorem}

\newtheorem{lemma}[theorem]{Lemma}

\newtheorem{definition}{Definition}

\DeclareMathOperator{\median}{median}
\DeclareMathOperator{\rank}{rank}
\DeclareMathOperator{\cnt}{count}
\DeclareMathOperator{\Lap}{Lap}
\DeclareMathOperator{\Var}{Var}
\DeclareMathOperator{\Err}{Err}
\DeclareMathOperator{\anc}{anc}
\DeclareMathOperator{\desc}{desc}
\DeclareMathOperator{\child}{child}
\DeclareMathOperator{\parent}{par}
\DeclareMathOperator{\cov}{cov}
\DeclareMathOperator{\diag}{diag}
\renewcommand{\SS}{\mathop{\mathrm{SS}}}

\DeclareMathOperator{\EM}{EM}

\DeclareMathAlphabet{\mathcal}{OMS}{cmsy}{m}{n}

\newcommand{\reals}{{\mathbb R}}
\newcommand{\data}{{\mathbb D}}
\newcommand{\eat}[1]{}
\newcommand{\A}{{\cal A}}
\newcommand{\D}{{\cal D}}
\newcommand{\T}{{\cal T}}
\renewcommand{\L}{{\cal L}}
\newcommand{\eps}{{\varepsilon}}
\newcommand{\qed}{\hfill \rule{1ex}{1ex}}
\title{Differentially Private Spatial Decompositions}
\author{Graham Cormode, Magda Procopiuc, Divesh Srivastava\\
AT\&T Labs--Research\\
\{graham,magda,divesh\}@research.att.com
 \and Entong Shen, Ting Yu\\
 North Carolina State University\\
 \{eshen,tyu\}@ncsu.edu}
\date{}
\begin{document}
\maketitle

\begin{abstract}
Differential privacy has recently emerged as the de facto standard for private
data release.
This makes it possible to provide
strong theoretical guarantees on the privacy and utility of released
data.
While it is well-understood how to release data based on counts and simple
functions under this guarantee, it remains to provide general purpose
techniques that are useful for a wider variety of queries.
In this paper, we focus on spatial data, i.e., any multi-dimensional
data that can be indexed by a tree structure.
Directly applying existing differential privacy methods to this type
of data simply generates noise.

We propose instead the class of ``private spatial decompositions'':
these adapt standard spatial indexing methods such
as quadtrees and kd-trees to provide a private description of the
data distribution.
Equipping such structures with differential privacy requires several
steps to ensure that they provide meaningful privacy guarantees.
Various basic steps, such as choosing splitting points and describing
the distribution of points within a region, must be done privately,
and the guarantees of the different building blocks must be composed into
an overall guarantee.
Consequently, we expose the design space for private spatial
decompositions, and analyze some key examples.
A major contribution of our work is to
provide new techniques for parameter setting and
post-processing of the output to improve
the accuracy of query answers.
Our experimental study demonstrates that it is possible to build such
decompositions efficiently, and use them to answer a variety of
queries privately and with high accuracy.
\end{abstract}

\section{Introduction}
\label{sec:intro}

Releasing representative data sets which do not
compromise the privacy of data subjects has occupied serious effort in
the database community in recent years.
The paradigm of differential privacy has recently emerged as the
favored definition: it ensures that what can be learned from
the released data does not substantially differ whether or not any
given individual's data is included.
This is intended to reassure data subjects that their participation in the
process does not directly lead to information about them being
revealed.
The question that the data management community must now address is how
to provide differential privacy guarantees, while ensuring that the result
is not just private, but also useful.

Consider a dataset containing locations of individuals
at a particular time, e.g., GPS locations or home addresses.
Such data could be used for many applications:
transportation planning, facility location, political boundary drawing
etc. More generally, any data set where attributes are ordered and have
moderate to high cardinality (e.g., numerical attributes such as salary)
can be considered spatial data: whenever data can be indexed by a
tree structure (such as a B-tree, R-tree, kd-tree etc.), it is
implicitly being treated as spatial.
In spatial applications, a basic primitive is to know how many individuals
fall within a given region (a multi-dimensional range query).
Our aim is to release information which allows such queries to be
answered accurately, while giving a strong guarantee of privacy on
people's locations.

Although this is a natural and important problem, there has been
limited prior work that can be applied directly to this setting.
The most straightforward method is to lay down a fine grid over the data, and
add noise from a suitable distribution
to the count of individuals within each cell~\cite{Dwork:06}.
For example, suppose we represent a set of $10^7$ GPS locations
in a grid of 10 meter $\times$ 10 meter squares over the USA territory:
this yields a total of approximately $10^{11}$ entries, most of which
are either 0 or 1. Because the purpose of the noise is to mask whether there is an
individual there, the output is simply a large mass of
noisy counts, with little information remaining to accurately answer
queries. Any query which touches, e.g., a 1\% fraction of the area
includes over $10^9$ noisy counts, which translates into a huge error.

In this paper, we aim to balance the requirements of practicality and
utility, while achieving a desired privacy guarantee.
In particular, we design a class of differentially private
spatial decompositions (PSDs).
These partition the space into smaller regions, and report statistics
on the points within each region.
Queries can then be answered by intersecting the query region with the
decomposition.
\eat{
The motivation is that privacy mechanisms work poorly when the count is very small
(because the noise dominates the count value),
 and query answering is inaccurate when the region is very large
 (because of skewness in data distribution).
}
By performing a spatial decomposition which results in compact
regions with sufficiently many points and a more uniform distribution,
we expect query answers to be more accurate.

There are many existing (non-private) approaches to spatial decompositions.
Some are {\em data-independent}, such as quadtrees
which recursively divide the data space into equal quadrants.
Other methods, such as the popular kd-trees, aim to better capture the data
distribution and so are {\em data-dependent.}
In this case, there are additional challenges:
the description of the regions must also be differentially private.
In other words, simply building a tree-structure over the exact data
and populating it with noisy counts is not sufficient to meet the
differential privacy definition: the choice of splits is
based on the true data, and potentially reveals information.

A second issue is how to answer queries given a PSD.
In the non-private case, it is fairly straightforward to take a query
region and intersect it with the data structure to obtain a count of
contained nodes.
In the private case, it is more complicated: because we have count
estimates for large regions and subregions, there are multiple ways to
decompose a query region, and the answers may vary due to differing
noise.
\eat{
In this work, we show how to build optimal estimates from the multiple
views of the data.
In analysis of query accuracy, we provide upper bounds on the number
of nodes intersecting an arbitrary range query.
Based on this, we propose several budget allocation schemes and
theoretically compare their implications on overall noise variance.
}
To achieve optimal query accuracy, we develop two new techniques that
may be of independent interest:

\begin{itemize}
\item First, we show that using non-uniform
noise parameters can significantly improve accuracy, while
maintaining the same privacy guarantee. Specifically, we propose
setting the parameters in a geometric progression, increasing from
root to leaves. To the best of our knowledge, this is the first result
that analyzes the impact of non-uniform noise parameters.

\item Second, we design a new method to compute optimal (minimum variance)
answers to queries by post-processing the noisy counts. Our method
works over a large class of non-uniform noise parameters. It generalizes
the method of~\cite{Hay:Rastogi:Miklau:Suciu:10}, which applied only to uniform
noise parameters.
\end{itemize}

Putting these pieces together,
we show how to build private versions of well-known data structures,
such as kd-trees, R-trees, and quadtrees, in multiple
dimensions.
This gives a full framework for privately representing spatial data.
Our experimental study ranges over this framework, and allows us to
see the impact of different choices on the accuracy of query answering
over a mixture of real and synthetic data. We show that our two novel techniques
can reduce the absolute error of queries by up to an order of magnitude.

\para{Outline.}
In Section~\ref{sec:related} we describe related work, while
Section~\ref{sec:prelim} surveys concepts from differential privacy
and spatial decompositions.
We build our framework as follows:
In Section~\ref{sec:counts},
we show how to set non-uniform noise parameters in a hierarchical structure.
Section~\ref{sec:postprocess} describes the post-processing method.
%
In Section~\ref{sec:ddtrees},
we describe how to effectively find 
private median points under
differential privacy, with further extensions in
 Section~\ref{sec:improve}.
Our experimental study in Section~\ref{sec:experim}
considers different optimizations for different families of PSDs, and
then compares these all together to find the best
choices for working privately with spatial data.
We also compare to two prior
approaches \cite{Xiao:Xiong:Yuan:10,Inan:2010:PRM}
and show that our methods outperform them.

\section{Related Work}
\label{sec:related}
In this section, we present a brief summary of the most related topics
in anonymization and privacy.  For more details, there are several
recent surveys and tutorials
\cite{Chen:Kifer:Lefevre:Machanavajjhala:10,Cormode:Srivastava:09,Dwork:11,Gehrke:Kifer:Machanavajjhala:10}.

Initial efforts to ensure privacy of released data were
based on syntactic definitions
such as $k$-anonymity~\cite{Samarati:Sweeney:98} and
$\ell$-diversity~\cite{Machanavajjhala:Gehrke:Kifer:Venkitasubramaniam:06}.
Subsequent efforts tried to provide a more semantic guarantee,
ensuring that no matter what knowledge or power an adversary had,
their ability to draw conclusions about an individual in the data was
limited.
This culminated in the definition of differential privacy~\cite{Dwork:06}
which ensures that the probability of any property holding on the
output is approximately the same, whether or not an individual is
present in the source data.

Although often introduced in the context of an interaction between a
data owner and a data user, differential privacy naturally adapts to a
non-interactive setting where the data owner wishes to release private
information derived from their data.
Conceptually, this is quite straightforward: the data owner determines
what collection of statistics should be released, then computes these
under some mechanism which provides an overall privacy guarantee.
Initial work showed how to release contingency tables
(equivalent to count cubes) in this `publishing'
model~\cite{Barak:Chaudhuri:Dwork:Kale:McSherry:Talwar:07}.
Subsequent work in the database community also adopted the model
of releasing tables of counts (histograms) and studied how to
ensure these are accurate for different query workloads \cite{Li:Hay:Rastogi:Miklau:McGregor:10,Xiao:Wang:Gehrke:10}.

There has been much work in producing mechanisms which
offer differential privacy, such as
 the general purpose exponential mechanism \cite{McSherry:Talwar:07},
and the geometric mechanism \cite{Ghosh:Roughgarden:Sundararajan:09}.
Approaches such as smoothed sensitivity have attempted to adapt the
amount of noise needed to the instance of the data \cite{Nissim:Raskhodnikova:Smith:07}.
These mechanisms have been applied to various problems, such as
releasing contingency tables
\cite{Barak:Chaudhuri:Dwork:Kale:McSherry:Talwar:07},
time series \cite{Rastogi:Nath:10}, and
recommender systems \cite{McSherry:Mironov:09}.
Recently, efforts have been made to improve the accuracy of (count)
queries, by allowing ranges to be expressed via a smaller number of
noisy wavelet or range coefficients
\cite{Li:Hay:Rastogi:Miklau:McGregor:10,Xiao:Wang:Gehrke:10}, and via
post-processing with overlapping information
\cite{Hay:Rastogi:Miklau:Suciu:10}.

However, there has been limited work that applies to spatial data.
The idea of differentially private
data-partitioning index structures was previously
suggested in the context of private record
matching~\cite{Inan:2010:PRM}.
The approach there is based on using an approximate mean as a
surrogate for median (on numerical data) to build kd-trees.
This becomes a special case in our general framework.
In Section \ref{exp:median}, we show that there are much better
methods to choose medians privately.
We also compare to this case as part of our overall experimental
study, and also for a record matching application in
Section~\ref{exp:edbt}.

The only prior work directly addressing spatial data
follows the approach suggested above,
and imposes a fixed resolution grid over the base
data~\cite{Xiao:Xiong:Yuan:10}.
It then builds a kd-tree based on noisy counts in the grid,
splitting nodes which are not considered ``uniform'',
and then populates the final leaves with ``fresh'' noisy estimated counts.
In our experimental study (Section~\ref{sec:experim}), we observe
that this approach is inferior to other points in the framework.

\section{Preliminaries}
\label{sec:prelim}

\subsection{Differential Privacy}\label{defs}
We now formally introduce the concepts behind differential privacy.
Let $\D_1,\D_2$ be two neighboring datasets, i.e., $\D_1$ and $\D_2$
differ in only one tuple $t$, written as $\|\D_1 - \D_2\|=1$.
In some cases, this is taken to mean
that $t$ has different values in the two datasets;
in other cases, it is interpreted as meaning
that $t$ is present in only one of the two datasets.
Either version leads to privacy guarantees which differ by at most
 a constant factor.
For concreteness, in this paper we consistently use the second
definition.

\begin{definition}\label{diffpriv}
Let $\D_1,\D_2$ be two neighboring datasets.
Let $\A$ denote a randomized algorithm over datasets,
and $S$ be an arbitrary set of possible outputs of $\A.$ 
%
Algorithm $\A$ is said to be $\eps$-differentially private if
$$\Pr[\A(\D_1)\in S]\leq e^{\eps}\Pr[\A(\D_2)\in S].$$
%
%
\end{definition}

\smallskip
Intuitively, differential privacy guarantees that no individual tuple
can significantly affect the released information:
the output distribution generated by $\A$ is nearly the same, whether or
not that tuple is present in the dataset. The most common technique for
designing differentially-private algorithms was proposed
in~\cite{Dwork:McSherry:Nissim:Smith:06}. It is a noise adding mechanism,
as follows.

\begin{definition}\label{laplace}
{\bf (Laplace mechanism)}
Let $f(\D)$ denote a numeric function over dataset $\D.$
An $\eps$-differentially private mechanism for releasing $f$ is to publish
$\L(\D) = f(\D) + X,$ where $X$ is a
random variable drawn from the Laplace distribution $\Lap(\sigma(f)/\eps)$.

The value $\sigma(f)$, called the {\em sensitivity of $f,$} is the maximum change in $f$
when any single tuple of $\D$ changes. Formally:
$$\sigma(f) = \max_{\D_1,\D_2: \|\D_1-\D_2\| = 1 } |f(\D_1)-f(\D_2)|.$$
\end{definition}

For example, if $f= \operatorname{count}$, then $\sigma(f) = 1:$
for any two neighboring datasets $\D_1$ and $\D_2$,
the cardinalities of $\D_1$ and $\D_2$ differ by 1.

\eat{
The following result was proven
in~\cite{Dwork:McSherry:Nissim:Smith:06}.
A similar result was shown in~\cite{Ghosh:Roughgarden:Sundararajan:09}
for the geometric distribution, instead of Laplace.

\begin{lemma}[\cite{Dwork:McSherry:Nissim:Smith:06}]
\label{noise}
Let $\eps >0$ and let $f: \data \rightarrow \reals$ be a function with
sensitivity $\sigma(f).$
If $X$ is a random variable drawn from the Laplace distribution with parameter
$\sigma(f)/\eps$, i.e. $X \sim \Lap(\sigma(f)/\eps))$,
then the randomized algorithm
$\A(\D) = f(\D) + X$ is $\eps$-differentially private.
\end{lemma}
}

\subsection{Spatial Decompositions}
\label{subsec:trees}
Spatial indexes are routinely used in database management, with a long history stretching
over decades~\cite{Samet:06,deBerg:Cheong:Kreveld:Overmars:08}. Their main purpose is
to allow queries to determine quickly whether or not a particular
point is present in the data set. Clearly, we cannot extend this functionality to a privacy-preserving
setting.
However, spatial indexes also allow efficient computation for a rich set of aggregate queries,
particularly {\em range queries}.
We focus on these queries when studying the utility of PSDs.

Formally, a {\em spatial decomposition} is a hierarchical
(tree) decomposition of a geometric
space into smaller areas, with data points partitioned among the
leaves.
Indexes are usually computed down to a level where the leaves either
contain a small number of points,
or have a small enough area, or a combination of the two.
Unless otherwise specified, we assume
that the tree is complete, i.e., all leaf-to-root paths have the same
length, and all internal nodes have the same fanout.

\begin{figure}[t]
\centering
\includegraphics[width=0.5\columnwidth]{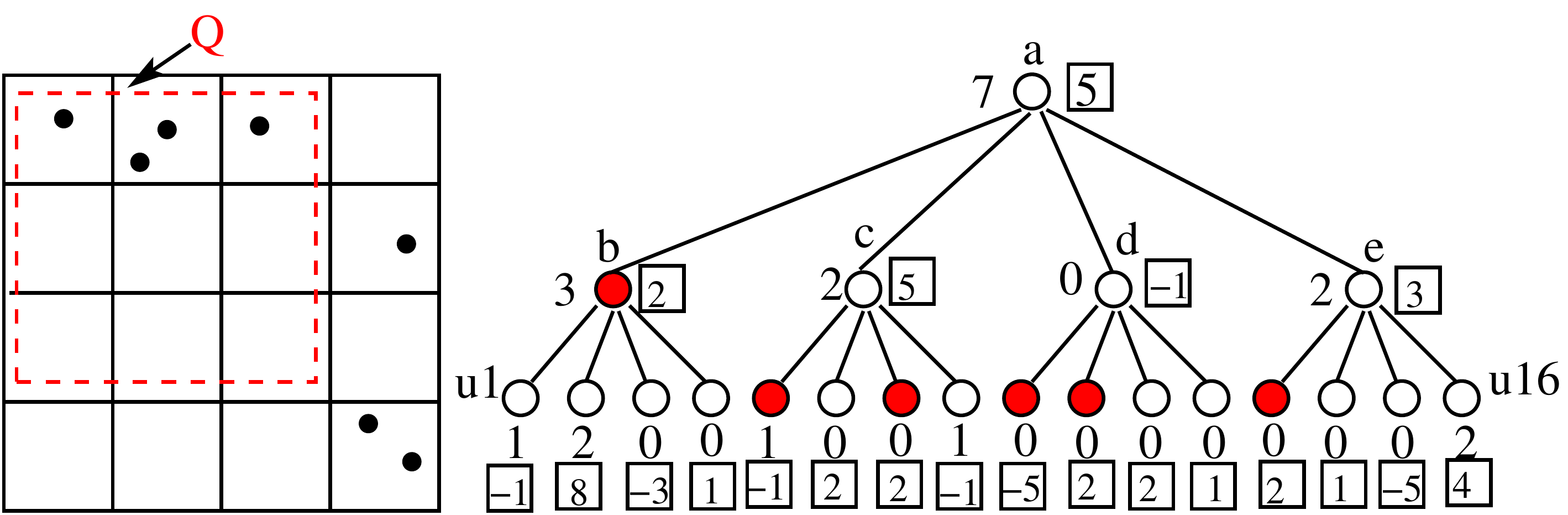}
\caption{Example of private quadtree: noisy counts (inside boxes) are released; actual counts, although
depicted, are not released.
Query $Q$ (dotted red rectangle) could be answered by adding noisy counts of marked nodes.}
\label{fig:toyquadtree}
\end{figure}

%
%

The following classification of spatial decompositions is widely used.
We will describe our private spatial decompositions following the same taxonomy.

\para{Data-independent decompositions.}
The best known example in this category is
the {\em quadtree} in two dimensions, and its generalizations to higher
dimensions (octree, etc.)
Their defining feature is that their
structure is pre-defined, and depends only on the domain of the
attributes.
E.g., in the quadtree, nodes are recursively divided into four equal regions via horizontal and vertical lines through
the midpoint of each range~\cite{deBerg:Cheong:Kreveld:Overmars:08}.

\eat{
\begin{figure}[t]
\centering
\begin{tabular}{cc}
\begin{minipage}{0.3\columnwidth}
\centering
\footnotesize
\includegraphics[width=10mm]{hilbert} \\

(a) \\

\includegraphics[width=20mm]{hilbert2}

(b)
\end{minipage}
&
\begin{minipage}{0.5\columnwidth}
\centering
\footnotesize
\includegraphics[width=32mm]{hilbert3} \\

(c)
\end{minipage}
\end{tabular}
\caption{Hilbert curves of order (a) 1 (b) 2 and (c) 3}
\label{fig:hilbert}
\end{figure}
}

\para{Data-dependent decompositions.} Here, the hierarchical partition
of nodes is based on the input. Examples include:

\noindent
{\em -- Kd-trees:} Nodes in a kd-tree are recursively split via lines passing through the median data value along some coordinate axis.
The splitting axis changes at each level, in a cyclic order.

\noindent
{\em -- Hilbert R-trees:} The R-tree is a spatial decomposition formed by
 nested (hyper)rectangles which may overlap.
A Hilbert R-tree is computed as follows:
  Map all points to a Hilbert space-filling curve of sufficiently
  large order (see~\cite{Kamel:Faloutsos:94} for details).
Build a binary tree on the Hilbert values, then map
it back into the original space.
The bounding boxes of data
points corresponding to nodes in the tree form the Hilbert R-tree.

\para{Hybrid trees.}
Although less studied in the classical setting,
a hybrid approach can be particularly useful in a privacy-aware model.
In particular, we consider using  data-dependent splits for the first
$\ell$ levels of the tree (for some $\ell$ decided in advance),
then switch to data-independent splitting (e.g., quadtree)  for the
remainder.
We observe in Section~\ref{exp:query} that queries over private
hybrid trees are generally more accurate than with many other indexes.

\subsection{Building Private Spatial Decompositions}\label{subsec:privatetrees}

We are now ready to describe how to build PSDs.
The simplest PSD is formed by instantiating a data-independent tree
such as a full quadtree, but computing
counts for each node via the Laplace mechanism. See Figure~\ref{fig:toyquadtree}
for an example.

Recall that the concept of sensitivity (Definition~\ref{laplace})
captures the maximum change in the output caused by the presence or
absence of a single tuple.
For {\em data-independent} trees, releasing the structure of the index
(i.e., the node rectangles) does not endanger the privacy of any
individual.
The only data-dependent features are the node counts: each node stores
the number of data points that lie in the spatial cell associated with
it.
Note that adding or deleting a single tuple changes the counts of all
the nodes on the path from the root to the leaf containing that tuple.
Therefore, to obtain a tree that satisfies $\eps$-differential privacy
as a whole,
we need to combine the privacy guarantees of individual node counts.
We use the following well-known composition lemma:

\eat{
\begin{lemma}~\cite{McSherry:Mironov:09}\label{epscomp}
Let $\A_1,\ldots,\A_t$ be $t$ algorithms such that $\A_i$ satisfies
$(\eps_i,\delta_i)$-differential privacy, $1\leq i\leq t.$ Then their
sequential composition $\langle\A_1,\ldots,\A_t\rangle$ satisfies
$(\eps,\delta)$-differential privacy, for
$\eps = \sum_{i=1}^t\eps_i$ and $\delta = \sum_{i=1}^t \delta_i$.
\end{lemma}
}
\begin{lemma}~\cite{McSherry:Mironov:09}\label{epscomp}
Let $\A_1,\ldots,\A_t$ be $t$ algorithms such that $\A_i$ satisfies
$\eps_i$-differential privacy, $1\leq i\leq t.$ Then their
sequential composition $\langle\A_1,\ldots,\A_t\rangle$ satisfies
$\eps$-differential privacy, for
$\eps = \sum_{i=1}^t\eps_i$.
\end{lemma}

\smallskip
In any partition tree, we only need to consider sequential compositions
of private counts along a path:
if nodes $u$ and $v$ are not on the
same root-to-leaf path, their node counts are independent of each other, so
knowing the output of $\A_u$ does not affect the privacy guarantee of $\A_v.$

This outline provides a simple PSD.  However, 
we have developed several optimization techniques which can
significantly improve the utility of PSDs, which is the technical
focus of this paper.
Section~\ref{sec:counts} shows how to choose the values $\eps_i$,
while Section~\ref{sec:postprocess} shows how to process the noisy counts to
maximize query accuracy.
These results apply to all spatial decompositions.
For data-dependent and hybrid trees,
to guarantee differential privacy,
the structure of the tree itself must also be perturbed by noise.
We study this issue in
Section~\ref{sec:ddtrees} for kd-trees.
We also consider building private Hilbert R-trees, which are binary
trees (i.e., one-dimensional kd-trees) in the Hilbert space.
When mapping back to the original space, we use the bounding boxes of
Hilbert values corresponding to tree nodes to define the node rectangles (thus preserving privacy).
For brevity, we do not explicitly discuss Hilbert R-trees further,
although we include them in our experimental evaluation
in Section~\ref{sec:experim}.

\section{Allocating Noise Parameters}
\label{sec:counts}

In this section, we focus on how to choose noise parameters $\eps_i$
such that the composition rule is satisfied
for all tree paths (see Section~\ref{subsec:privatetrees}). Let $h$ denote the height of the tree; leaves have level 0 and the root
has level $h$. We assume that all nodes at level $i$ have the same Laplace parameter $\eps_i$ (other choices are
discussed at the end of this section).
Hence, given a total privacy ``budget'' of $\eps$, we need to
specify $\eps_i$, $0\leq i\leq h$, such that
$\sum_{i=0}^h\eps_i = \eps$. We refer to a choice of $\eps_i$'s as a {\em budget strategy}. The goal is to minimize the resulting query errors.


\para{Error measure.} For any query $Q$, let $\tilde{Q}$ denote the answer to
$Q$ computed over the private tree.
Then $\tilde{Q}$ is a random variable which is an unbiased estimator of the true answer (since noise
has mean 0).
Its variance $\Var(\tilde{Q})$ is a
strong indicator of query accuracy.
As in prior work~\cite{Xiao:Wang:Gehrke:10,Li:Hay:Rastogi:Miklau:McGregor:10},
we define the error measure to be $\Err(Q) = \Var(\tilde{Q})$.
The error of a query workload
$Q_1,\ldots,Q_s$ is $\sum_{i=1}^s \Err(Q_i) / s.$

\subsection{Query Processing}\label{sec:qa}

Unlike a tree over the original (unperturbed) counts, a PSD
may return many different results to a query $Q$.

\para{Query processing example.}
Figure~\ref{fig:toyquadtree} shows a possible processing for query $Q$, which sums the noisy counts
in nodes $b$, $u5$, $u7$, $u9$, $u10$ and $u13$. The answer is $2$. However, if we replace
$b$'s count by the sum of its children's counts, and the sum of $u5$ and $u7$ by the difference
between $c$'s count and the sum of $u6$, $u8$, the answer becomes $8$.
This is because the noise is independent, and there are multiple ways of representing $Q$ as a union
or difference of node rectangles. Adding/subtracting corresponding noisy counts yields
different results.
\qed

\smallskip
Consequently, to analyze $\Err(Q)$, we must first
describe a standard method for computing $\tilde{Q}$.
Let $Y$ be the set of  noisy counts, and
let $U$ be the set of nodes used to answer $Q$.
Then $\Err(Q) = \sum_{u \in U} \Var(Y_u)$, i.e., the
total variance is the sum of the node variances.
Thus the error grows, to a first approximation, with
the number of noisy counts included.
We adopt the canonical range query processing
method~\cite{deBerg:Cheong:Kreveld:Overmars:08}, which minimizes
the number of added counts.

\eat{
We therefore adopt the canonical
method~\cite{deBerg:Cheong:Kreveld:Overmars:08} for processing $Q$,
which is guaranteed to use at most a constant number of counts from
each level of the tree
to answer $Q$.
}

The method is as follows: Starting from the root, visit all nodes $u$
whose corresponding rectangle intersects $Q$.
If $u$ is fully contained in $Q$, add the noisy count $Y_u$
to the answer $\tilde{Q}$; otherwise,
recurse on the children of $u,$ until the leaves
are reached.
If a leaf $a$ intersects $Q$ but is not contained in $Q$, use a {\em
  uniformity assumption} to estimate what fraction of $Y_a$ should be
added to $\tilde{Q}$.

Let $n(Q)$ be the number of nodes that contribute their counts to $\tilde{Q}$.
For each $0\leq i\leq h$,
let $n_i$ be the number of nodes at level $i$ that are maximally contained in $Q$
(i.e. contributed their counts to $\tilde{Q}$ according to the method
above), so $n(Q) = \sum_{i=0}^h n_i$.
The following result bounds
each $n_i$ and will guide us in choosing noise parameters $\eps_i.$


\eat{
We show that for all two-dimensional spatial decompositions we consider in  of height $h$ and fanout $f$,
$n(Q)=O(f^{h/2})$; note that the number of leaves in such a tree is $f^{h}$. In particular, for
the quadtree, $f=4$ so $n(Q) = O(4^{h/2}).$
}

\begin{lemma}\label{querybound}
Let $Q$ be an arbitrary range query,
and $\T$ be a spatial decomposition of height $h$ in two dimensions.
 Then

(i) If $\T$ is a quadtree, $n_i\leq 8\cdot 2^{h-i}$ and
 \[\textstyle n(Q) \leq 8(2^{h+1}-1) = O(4^{h/2}).\]\
(ii) If $\T$ is a kd-tree, $n_i\leq 8\cdot 2^{\lfloor (h-i+1)/2\rfloor}$ and
\[\textstyle n(Q) \leq 8(2^{\lfloor (h+1)/2\rfloor+1}-1) = O(2^{h/2}).\]

\end{lemma}
\noindent
\textit{Proof.} 
Let $f$ denote the fanout of $\T$: $f=4$ for the quadtree and $f=2$ for the kd-tree.
Consider the left vertical extent of $Q$. If $\T$ is a quadtree, then
for each rectangle at level $i$ intersected by this vertical
line, there are 2 rectangles at level $i-1$ intersected by it.
If $\T$ is a kd-tree, then for each rectangle that the vertical extent intersects at level $i$
(assume wlog that level $i$ corresponds to a vertical split), it intersects 1 at
level $i-1$ and 2 at level $i-2$. In either case,
the number of rectangles intersected by the left extent of $Q$ grows by at most a factor $f$
every two levels.

We repeat this argument for the other three extents of $Q$.
Additionally, the number of nodes that are maximally contained in $Q$ can be
bounded by the number of nodes that are intersected by $Q$, and the
recurrence for $n_i$ follows.
\qed

\medskip

We remark that a similar result extends to $d$ dimensional decompositions, where
the behavior is $n(Q) = O(f^{h(1-1/d)})$ for a tree of height $h$ and fanout $f$.

\subsection{Budget Strategies}

The variance of the Laplace mechanism with parameter $\eps_i$
is $\Var(\Lap(\eps_i))= 2/\eps_i^2$.
Since the noise is independently generated in each node, we deduce
 that
\begin{equation}\textstyle \Err(Q) = \sum_{i=0}^h 2n_i/\eps_i^2.
\label{eq:var}
\end{equation}
For simplicity, we analyze strategies only for quadtrees; hence, $n_i \leq 8 \cdot 2^{h-i}$ by Lemma~\ref{querybound}(i).
The analysis for kd-trees is similar.

\para{Uniform Budget:}
A natural strategy is to set $\eps_i=\eps/(h+1).$
Prior work that finds counts in  trees
(e.g. \cite{Hay:Rastogi:Miklau:Suciu:10}) has used this model.
Then
$\Err(Q)$ $=$ $\frac{2(h+1)^2}{\eps^2}\sum_{i=0}^h n_i$ $\leq$ $\frac{16}{\eps^2}(h+1)^2(2^{h+1}-1).$

\para{Geometric Budget:}
We can significantly improve the query accuracy by
considering a non-uniform budgeting strategy.
Substituting the bound for $n_i$ into \eqref{eq:var}
we obtain the following optimization problem to minimize the upper
bound.
$$\begin{array}{ll}
\text{Minimize} & \sum_{i=0}^h 2^{h-i}/\eps_i^2\\
\text{Subject to:} & \sum_{i=0}^h\eps_i = \eps.
\end{array}$$
\begin{lemma}\label{lem:geobudget}
An upper bound for $\Err(Q)$ is
$$\frac{16(2^{(h+1)/3}-1)^3}{\eps^2(\sqrt[3]{2}-1)^{3}} \leq \frac{2^{h+7}}{\eps^2}$$
which is attained for
$\eps_i =2^{(h-i)/3}\eps\frac{\sqrt[3]{2}-1}{2^{(h+1)/3}-1}$.
\end{lemma}

\noindent
\textit{Proof.}
By the Cauchy-Schwarz inequality, we have
$$(\sum_{i=0}^h\eps_i)(\sum_{i=0}^h 2^{h-i}/\eps_i^2)\geq
         (\sum_{i=0}^h \sqrt{\eps_i 2^{h-i}/\eps_i^2})^2,$$
with equality attained if and only if $\eps_i = C2^{h-i}/\eps_i^2,\forall i,$
where $C$ is a constant. Hence,
$\eps_i=\sqrt[3]{C}2^{(h-i)/3}.$
Plugging this into $\sum_{i=0}^h\eps_i = \eps,$ we obtain
$\sqrt[3]{C}=\frac{\eps (\sqrt[3]{2}-1)}{2^{(h+1)/3}-1}$.
Then $\Err(Q)$ is at most	
\[16\sum_{i=0}^{h} 2^{h-i}/(C^{2/3}2^{2(h-i)/3}) = 16C^{-2/3} \sum_{i=0}^h 2^{(h-i)/3} =
16\eps/C,\] and the result follows.
\qed

\medskip
Lemma~\ref{lem:geobudget} shows that query accuracy is improved
with a geometric budgeting scheme: starting from the root, the budget
increases geometrically (by a factor of $2^{1/3}$), so the leaf counts
are reported with highest accuracy.

Studying the bound in the lemma,
it seems that we should reduce $h$ to reduce the noise.
But this variance only bounds the error from noisy counts.
We also have to account for the error due to queries which partly
intersect some leaves, i.e., errors arising from the uniformity assumption.
In the worst case, this error is proportional to the number of points
in each leaf intersected by the query.
We intersect $n_h\approx 2^h$ leaves. On the average for an input of
$n$ points, we could
have $O(n/4^h)$ points per leaf (assuming balanced leaf counts).
Hence, the uniformity assumption error behaves as $O(2^h n/4^h) = O(n/2^h)$.
Then the overall error grows as $O(n/2^h + 2^{h/3})$, suggesting we
benefit more overall as $h$ increases.

\begin{figure}[t]
\centering
\includegraphics[width=0.4\columnwidth]{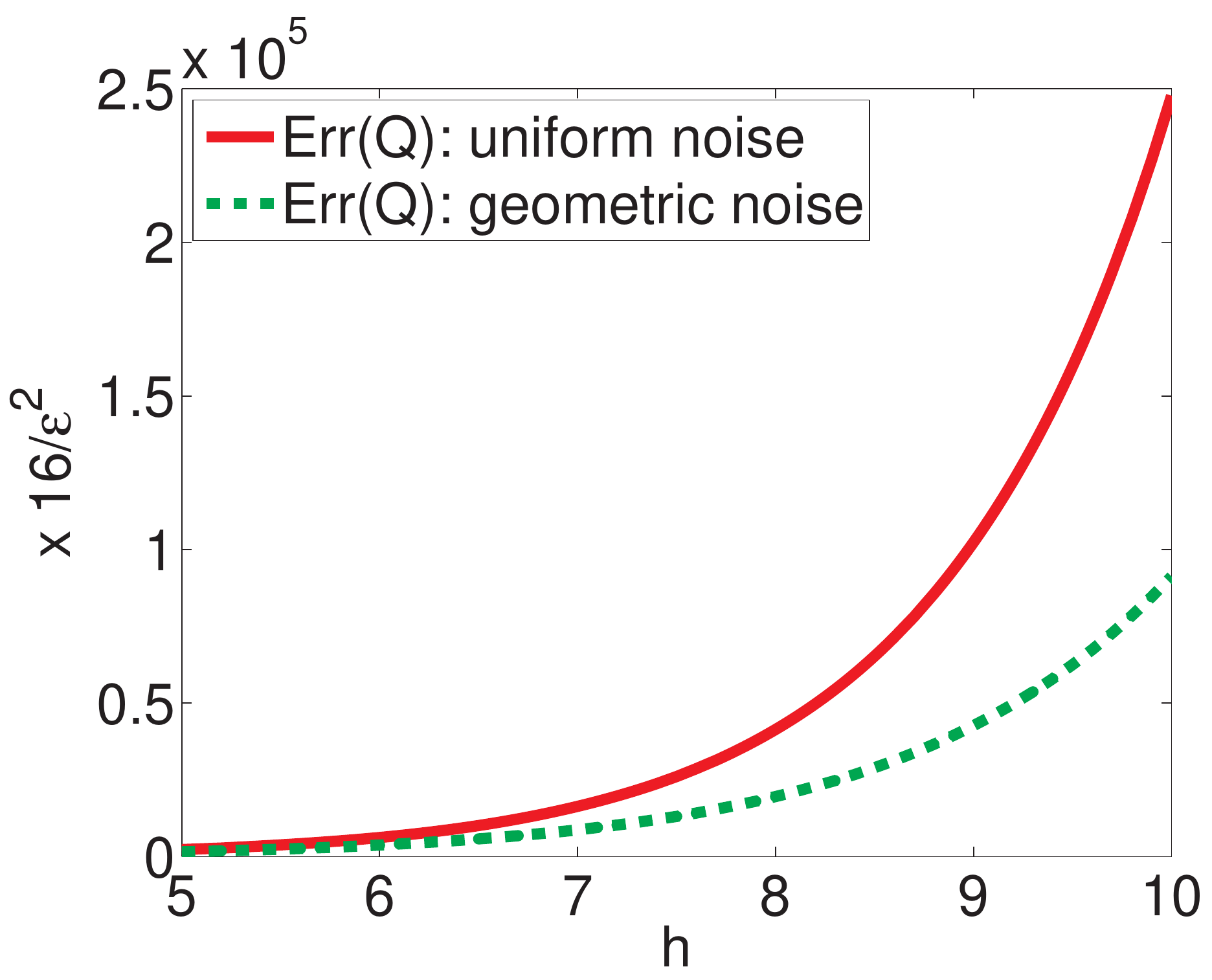}
\caption{Worst case behavior of $\Err(Q)$ for uniform and geometric noise.}
\label{fig:mathplot}
\end{figure}

\para{Comparing strategies.}
Let $Q$ be a query that includes the maximum number of counts at each
level, i.e., $n_i = 8\cdot 2^{h-i}.$\footnote{More rigorously, $n_i=\min\{8\cdot 2^{h-i}, 4^{h-i}\},$
but this does not alter the analysis significantly.}
\eat{
As shown above,
the uniform strategy achieves
$$\Err(Q) = \frac{16}{\eps^2}(h+1)^2(2^{h+1}-1),$$
while the geometric strategy achieves
$$\Err(Q) = \frac{16}{\eps^2}\frac{(2^{(h+1)/3}-1)^3}{(\sqrt[3]{2}-1)^{3}}.$$
}
Figure~\ref{fig:mathplot} shows that worst case error for
a uniform budget,
$\Err_{\operatorname{unif}}(h) = (h+1)^2(2^{h+1}-1)$ grows
much faster than the geometric budget error,
$\Err_{\operatorname{geom}}(h) =
  \frac{(2^{(h+1)/3}-1)^3}{(\sqrt[3]{2}-1)^{3}}.$
This comparison assumes that $Q$ touches a maximal number of nodes at
each level.
While in practice queries may touch fewer nodes,
our experiments show that uniform noise is still significantly less
accurate for a large number of different queries.

\para{Other budget strategies.}
There is a large class of strategies for dividing $\eps$ along a path.
For example, we could build a quadtree down to depth $h$ and
set $\eps_h = \eps,$ i.e., allocate the entire budget to leaves (this
approach is used in~\cite{Inan:2010:PRM}).
In this case, queries are computed over the grid
defined by the leaf regions and
the hierarchical structure of the tree is irrelevant.
Or we could conserve the budget by
setting $\eps_i=0$ for some levels $i$, and releasing no counts
for those levels. Queries then use counts from descendant nodes
instead. Conceptually, this is equivalent to increasing the
fanout of nodes in the tree.

Finally, we note that we do not have to use the same $\eps_i$ for all nodes on level $i$. For a generic query workload, this
approach makes sense, as each node on level $i$ is equally likely to be touched by the workload. However,  if the workload is
known a priori, one should analyze it to determine how frequently each node in the tree contributes to the answers. Then
$\eps_u$ could be larger for the nodes $u$ that contribute more frequently, subject to the sum along each path remaining $\eps.$

\eat{
{\em Leaf Budgeting:} We set $\eps_h = \eps_{\cnt},$ i.e., we allocate the entire budget along a path to the leaf. In this
case, queries are computed over the grid defined by the leaf regions.
The hierarchical structure of the tree is irrelevant.

{\em Skip Budgeting:} To optimize the use of budget $\eps_{\cnt},$ we set $\eps_i=0$
for some levels $i.$ We cannot release the accurate counts in those nodes, as this would violate differential
privacy. Therefore, we skip the counts altogether.
This means that instead of obtaining the count for some nodes
directly, a query computes the count from the sums of counts
of its descendants. For example, we set $\eps_i=0$ for odd $i$ in a
kd-tree, which is akin to flattening the kd-tree.
}

\section{Optimizing Query Accuracy}\label{sec:postprocess}

In the previous section we discussed how using geometric noise in a
hierarchical decomposition can significantly improve the accuracy of queries.
We now show that it is possible to further
improve query accuracy via a post-processing of the noisy counts.
The goal is to compute a new set of
counts for which query errors are minimized.
Note that this does not affect the privacy guarantee, as our method
takes as input the output of the differentially private mechanism.

\para{Post-processing example.}
Consider a simple tree with one root $a$ and four children, $b$, $c$,
$d$ and $e$.
Let $Y_v$ denote the noisy count of node $v\in\{a,b,c,d,e\},$ and
assume first that we used uniform noise of $\eps/2$ on each count.
There are two natural estimates of the true count of the root:
$Y_a$ itself, and the sum of counts of leaves, $Y_b + Y_c + Y_d +
Y_e$.
A first attempt is to estimate
$\beta_a = Y_a/2 + (Y_b+Y_c+Y_d+Y_e)/2$, i.e.,
the average of the two estimates.
In this case,
$\Var(\beta_a) = \Var(Y_a)/4 + 4\Var(Y_b)/4 = (5/4) \Var(Y_a),$
{\em worse} than directly using $Y_a$.
But it turns out we can do better:
setting $\beta_a = 4Y_a/5 + (Y_b+Y_c+Y_d+Y_e)/5$
yields $\Var(\beta_a) = (4/5) \Var(Y_a).$

For any non-uniform budgeting scheme, if the budget for $a$ is $\eps_1$ and
the budget of its children is $\eps_0$, then
$\beta_a = \frac{4\eps_1^2}{4\eps_1^2 + \eps_0^2} Y_a + \frac{\eps_0^2}{4\eps_1^2 + \eps_0^2}(Y_b+Y_c+Y_d+Y_e)$
improves accuracy:
one can derive
$\Var(\beta_a) = \frac{8}{4\eps_1^2 + \eps_0^2} < \frac{2}{\eps_1^2} = \Var(Y_a).$
\qed

\smallskip
As this example illustrates, there are many possible
ways of obtaining new counts as linear combinations of the published noisy counts.
The choices increase exponentially with the tree size, as we can combine counts
of ancestors, descendants, siblings etc. in various ways.
Luckily, we do not need to explore them in order to compute the best solution.
From linear statistical inference, we know that the best set of new
counts is obtained via the ordinary least-squares estimate (OLS)~\cite{Rao:65}.
In general, computing the OLS for $n$ unknowns
requires solving a linear system with $n \times n$ matrices.
Rather than explicitly inverting such (large) matrices,
we present a linear time algorithm that achieves the same result, by
taking advantage of the inherent symmetries of the matrices defined
for the tree structure.
Prior work considered the simpler case of uniform
noise parameters~\cite{Hay:Rastogi:Miklau:Suciu:10}.
We give a solution to the more general case when all nodes at level $i$
have the same Laplace parameter $\eps_i$
(this encompasses both uniform and geometric budgeting).
The generalization requires some technical effort to prove, but
yields a simple algorithm.

\para{Notation:}
As before, let $f$ denote the fanout of the spatial index and let $h$
denote its height.
We use $h(v)$ to denote the height of node $v$:
$h(v)=0$ if $v$ is a leaf, and $h(root) = h$.
We assume that the tree is complete, i.e., all paths have length $h$
and all internal nodes have fanout $f$.
Let $u \prec v$ denote that $u$ is a leaf in the subtree of $v$.
 Let $\anc(u)$ be the set of all ancestors of $u$, {\em including node
   $u$.}
We use $\parent(u)$ and $\child(u)$ to denote
the parent, resp. the set of children, of node $u.$

For linear inference over a tree structure, the
definition of an OLS is as follows.

\begin{definition}\label{def:minsq}
Let $Y$ denote the vector of original noisy counts, i.e., $Y_v$ is the
noisy count of node $v$.
Let $\eps_v$ denote
the noise parameter for node $v$. We denote by
$\beta$ the vector of counts after post-processing. Then $\beta$ is the {\em ordinary least squares estimator (OLS)}
if it is {\em consistent}, i.e.,
$\beta_v = \sum_{u\in \child(v)}\beta_u$ for all nodes $v$, and it minimizes $\sum_v \eps_v^2(Y_v - \beta_v)^2.$
\end{definition}

\smallskip

The OLS $\beta$ has two powerful properties:
It is
{\em unbiased} for any query $Q$ (recall that $Y$ is also
unbiased, since noise has mean 0).
Most importantly, among all
unbiased linear estimators derived from $Y$, it achieves
{\em minimum error for all range queries}.
In particular, it achieves smaller error than $Y$.

The computation of $\beta$ is based on the following result:

\begin{lemma}\label{lemma:rec}
For any node $v$ in the spatial index, the following recurrence holds
(with $Y$ and $\beta$ as in Definition~\ref{def:minsq}):
\begin{equation}
\!\!\!\!
\bigg(\sum_{j=0}^{h(v)} f^j \eps_j^2\bigg) \beta_v +
f^{h(v)} \!\!\!\!\!\!\!\!\! \sum_{w \in \anc(v)\setminus\{v\}}  \!\!\!\! \beta_w \eps_{h(w)}^2
= \sum_{u \prec v} \sum_{w \in \anc(u)}  \eps^2_{h(w)} Y_w
\label{eq:beta}
\end{equation}
\end{lemma}

\noindent
\textit{Proof.}
We use the notation defined in Section~\ref{sec:postprocess}. In addition, we define for each node $u$ the set
$\desc(u)$ to be the descendants of $u$, {\em including node $u$}. We
also use the shorthand $\anc(u,v) = \anc(u) \cap \anc(v).$
Let $C_u$ denote the true count of node $u$. We want to
show that if $\beta$ is the OLS for $C$, then it satisfies the recurrence in the statement of the lemma.
For any $v$
\begin{equation}
Y_v = {\sum_{u\prec v}} C_v + X_v
\label{eq:YC}
\end{equation}
where $X_v$ is the random variable drawn from the Laplace distribution
with parameter $\eps_{h(v)}.$
Therefore $\Var(Y_v) = 2/\eps^2_{h(v)}.$
The set of equations~\eqref{eq:YC} for all $v$ can be written in matrix form as
follows.
Let $n = f^h$ be the number of leaves, and let $m$ be the number of
nodes in the tree.
The binary matrix $H_{m\times n}$ representing the tree
hierarchy is defined as $H_{u,v} = 1$ if $v\prec u$ and 0 otherwise.
Assume wlog that the nodes $u$ are
in breadth-first order, and the leaves are ordered from left to
right.
Then \eqref{eq:YC} can be written as $Y = H\cdot C + X$, and the covariance matrix of $Y$ is $\cov(Y)=2\diag(1/\eps^2_{h(u)}).$

We apply the standard transformations
\[
 Z = (\cov(Y)/2)^{-\frac12} Y = \diag(\eps_{h(u)}) Y \text{ and }
U=(\cov(Y)/2)^{-\frac12} H
\]
to obtain the equation $Z$ $=$ $U\cdot C$ $+$ $\diag(\eps_{h(u)})\cdot X.$
Now $\cov(Z) = 2I$, where $I$ the unit matrix,
and the new equation fits the classical model.

Since the OLS is consistent, i.e., $\beta_v := \sum_{u \prec v} \beta_u,$  it suffices
to estimate the leaf counts, and the other estimates follow. A vector $\beta$ is the OLS for the
true count $C$ if it minimizes $(Z-U\beta)^T(Z-U\beta)$ (equivalent to Definition~\ref{def:minsq}). After differentiating, we have that $\beta$ satisfies
\begin{equation}
U^TU\beta = U^T Z
\label{eq:Z}
\end{equation}

By simple calculations, $(U^T U)_{u,w}$ $=$ $\sum_{v \in \anc(u,w)} \eps_{h(v)}^2$
and $(U^TZ)_u$ $=$ $\sum_{v \in \anc(u)} \eps_{h(v)} Z_v$ = $\sum_{v \in \anc(u)} \eps^2_{h(v)} Y_v.$
For any node $v$, we sum the corresponding rows on the left side of~\eqref{eq:Z} to obtain $\sum_{u \prec v} (U^T U)_u \beta =$

\begin{align*}
 &
\sum_{z \in [n]} \sum_{u \prec v} \sum_{w \in \anc(u,z) \setminus \anc(v)}
  \!\!\!\!\eps^2_{h(w)} \beta_z +
\sum_{z \in [n]} \sum_{u \prec v} \sum_{w \in \anc(u,z) \cap \anc(v)}
  \!\!\!\!\eps^2_{h(w)} \beta_z \\
= &
\sum_{z\prec v} \sum_{u \prec v} \sum_{w \in \anc(u,z) \setminus \anc(v)}
  \eps^2_{h(w)} \beta_z +
\sum_{w \in \anc(v)} \sum_{u \prec v} \sum_{z \prec w} \eps_{h(w)}^2 \beta_z
   \\
= &
\sum_{u \prec v} \sum_{w \in \anc(u) \setminus \anc(v)} \sum_{z \prec  w}
  \eps^2_{h(w)} \beta_z +
\sum_{w \in \anc(v)} \eps^2_{h(w)} \sum_{u \prec v}
   \left( \sum_{z \prec w} \beta_z \right) \\
= &
\sum_{u \prec v} \sum_{w \in \anc(u) \setminus \anc(v)} \eps^2_{h(w)}\beta_w
+ \sum_{w \in \anc(v)} f^{h(v)} \eps^2_{h(w)} \beta_w
\\ = &
\sum_{j=0}^{h(v)-1} \eps_j^2 \sum_{u \prec v} \sum_{w \in \anc(u) : h(w)=j} \beta_w
+ f^{h(v)} \sum_{w \in \anc(v)} \eps^2_{h(w)}\beta_w \\
= &
\sum_{j=0}^{h(v)-1} \eps_j^2 \sum_{w \in\desc(v) : h(w) = j} f^j \beta_w
+ f^{h(v)} \sum_{w \in \anc(v)} \eps^2_{h(w)}\beta_w \\
= &
\sum_{j=0}^{h(v)-1} \eps_j^2 f^j \beta_v
+ f^{h(v)} \eps^2_{h(v)}\beta_v
+ f^{h(v)} \sum_{w \in \anc(v)\setminus\{v\}} \eps^2_{h(w)}\beta_w\\
= &
\sum_{j=0}^{h(v)} \eps_j^2 f^j \beta_v
+ f^{h(v)} \sum_{w \in \anc(v)\setminus\{v\}} \eps^2_{h(w)}\beta_w
\end{align*}

We now sum the corresponding rows on the right hand side of~\eqref{eq:Z}, i.e., we sum
$(U^T Z)_u$ over $u \prec v$. Equating the left and right hand side, we
obtain~\eqref{eq:beta}.
\qed

\smallskip
Lemma~\ref{lemma:rec} provides an efficient way to compute $\beta$. First, pre-compute the following array $E$
of $h+1$ entries: $E_l = \sum_{j=0}^{l} f^j\eps^2_j.$ Since $E_l = E_{l-1} + f^l\eps^2_l$, this takes
time $O(h)$. For any node $v$, define $Z_v$ $=$ $\sum_{u \prec v} \sum_{w \in \anc(u)} \eps^2_{h(w)} Y_w.$
We compute $Z_v$ for all nodes $v$ in two linear traversals of the tree, as follows:

\smallskip
\noindent
 {\em Phase I: Top-down traversal} \\
We compute $Z_v$ for all leaves $v$. Note that in that case, $Z_v =  \sum_{w \in \anc(v)} \eps^2_{h(w)} Y_w.$
Let $\alpha_{root} = \eps^2_h Y_{root}$. In a top-down traversal of the tree, compute for each node $u$:
$\alpha_u$ $=$ $\alpha_{\parent(u)} + \eps^2_{h(u)} Y_u.$
When we reach a leaf $v$, we set $Z_v = \alpha_v$.

\smallskip
\noindent
{\em Phase II: Bottom-up traversal} \\
We compute $Z_v$ for all internal nodes $v$ as $Z_v$ $=$ $\sum_{u\in \child(v)} Z_u;$ this requires a single
bottom-up traversal.

\smallskip
\noindent
{\em Phase III: Top-down traversal}\\
We now compute $\beta_v$ for all nodes $v$. While doing so, we also compute an auxiliary value $F_v$, defined as
$F_v$ $=$ $\sum_{w \in \anc(v)\setminus\{v\}} \beta_w \eps_{h(w)}^2.$
Note that Equation~\eqref{eq:beta} for $v=root$ is:
\[ \bigg(\sum_{j=0}^{h} f^j \eps_j^2\bigg) \beta_{root}
(= E_h \beta_{root})
= Z_{root}\]

So we compute $\beta_{root} = {Z_{root}}/E_h$. In addition, let $F_{root} = 0.$
For any node $v\ne root$, assume we have already computed
$\beta_w$ and $F_w$ for all $w \in \anc(v)\setminus\{v\}$. Then we compute
$F_v$ $=$ $F_{\parent(v)} + \beta_{\parent(v)}\eps^2_{h(v)+1}.$
From Equation~\eqref{eq:beta}, we find
$$\beta_v = \frac{Z_v - f^{h(v)} \sum_{w \in \anc(v)\setminus\{v\}} \beta_w \eps_{h(w)}^2}{E_{h(v)}}
   = \frac{Z_v -  f^{h(v)}F_v}{E_{h(v)}}.$$

From Lemma~\ref{lemma:rec} and the above description we conclude:

\begin{theorem}
The algorithm consisting of Phases (I)--(III) described above computes the OLS estimator in time linear in
the size of the tree.
\end{theorem}

In Section~\ref{exp:query} we conduct an experimental evaluation which supports our theoretical result, and
shows significant improvement in query accuracy using the OLS.

\section{Data-Dependent and Hybrid Trees}\label{sec:ddtrees}

As noted in Section~\ref{subsec:trees}, publishing data-dependent
or hybrid trees with privacy guarantees must overcome an additional challenge:
the tree structure itself could reveal private information.
In most cases, the current node is split via a line through the
current median value along some axis.
For privacy reasons, we cannot reveal the exact median.
In Section~\ref{subsec:medians} we discuss methods for computing a
private median for a set of points.

Extending the computation of private medians to a
hierarchical structure requires some subtlety.
Note that, unlike private counts, it is no longer the case that only
root-to-leaf compositions matter.
If a tuple is deleted from the dataset,
it affects the (true) medians not only on its respective path,
but also for nodes in different parts of the tree.
To show the privacy of our approach,
we appeal to the fact that the composition
of differentially private outputs is well-understood in an interactive
model, where a user asks a series of queries.
We can imagine the following interaction:
At the root level, our private algorithm $\A$ outputs a noisy median
value $m_1.$
A user subsequently asks for the median of the points lying
on one side of $m_1,$ and for the median of the points lying on the
other side of $m_1$
(hence, $m_1$ becomes part of the user query). Algorithm $\A$
returns two noisy values, $m_2$ and $m_3$, each computed with respect
to the user-specified subset.
Hence, the computation can continue recursively, and it is
again sufficient to consider sequential compositions only along
root-to-leaf paths.
Now observe that
it is straightforward for the data owner to play both roles, and
output the final result (without interaction) as the PSD.

\subsection{Private Medians}\label{subsec:medians}

We now outline several methods for computing a private median for a
given set of points.
These approaches are either implicitly or explicitly developed in
prior work, but we are not aware of any previous comparison.
Here, we bring together the different approaches,
and give some new analysis on their behavior.
We  compare them empirically in Section~\ref{sec:experim}.

Let $C=\{x_1,$ $\ldots,$ $x_n\}$ be a (multi)set of $n$ values in
non-decreasing order in some domain range $[lo,hi]$ of size $hi-lo=M$,
and let $x_m$ be its median value.
We wish to compute a private median for $C$.
We could apply the Laplace mechanism (Definition~\ref{laplace}),
i.e., return $\L(C) = x_m + X$, where $X$ is Laplace noise.
However, the sensitivity of the median is of the same order of magnitude
as the range $M$
(see~\cite{Nissim:Raskhodnikova:Smith:07,Inan:2010:PRM}),
and the noise value $X$ usually dwarfs $x_m$.
A consequence of this is that  the noisy median is frequently outside
the range $[lo,hi]$.
When used in a kd-tree construction, such a noisy median does not help
to divide the data.
Instead, we study the following four methods.

\smallskip
\noindent
\textbf{Smooth Sensitivity}~\cite{Nissim:Raskhodnikova:Smith:07}
tailors the  noise to be more specific to the set $C$.
However, smooth sensitivity has slightly weaker privacy guarantees than
the Laplace mechanism: it only satisfies so-called $(\eps,\delta)$-differential
privacy; see~\cite{Nissim:Raskhodnikova:Smith:07} for details.

\begin{definition}(from Claim 3.3 of \cite{Nissim:Raskhodnikova:Smith:07})\label{smoothsens}
Let $0 < \eps,\delta < 1$, and let $\xi = \frac{\eps}{4(1+\ln(2/\delta))}.$
The {\em smooth sensitivity} of the median is defined as
$$\sigma_s(\median) = \max_{0\leq k\leq n} (e^{-k\xi}\max_{0\leq t\leq k+1}(x_{m+t} - x_{m+t-k-1})).$$
(where we define $x_i := lo$ if  $i < 0$ and $x_i$ $:=$ $hi$ if $i > n$).

The smooth sensitivity mechanism for median is defined as
$\SS(C) = x_m + \frac{2\sigma_s}{\eps}\cdot X$
where $X$ is a random variable drawn from the Laplace distribution
with parameter $1$ and $\sigma_s = \sigma_s(\median)$.
\end{definition}

\smallskip
\noindent
\textbf{Exponential Mechanism} is a general method proposed
in~\cite{McSherry:Talwar:07} as an alternative to the Laplace
mechanism: instead of adding random noise to the true value,
the output value is drawn from a probability distribution over all
possible outputs, so that the differential privacy condition is
satisfied.
Applied to median, we obtain the following algorithm:
\footnote{This computation is implicit in McSherry's PINQ
  system~\cite{McSherry:09}.}

\begin{definition}\label{expmech}
For any $x\in [lo,hi]$, let $\rank(x)$ denote the rank of $x$ in $C.$ The
exponential mechanism $\EM$ returns $x$ with
$\Pr[ \EM(C) = x ] \propto e^{-\frac{\eps}{2}|\rank(x) - \rank(x_m)|}.$
\end{definition}

Since all values $x$ between two consecutive values in $C$ have the
same rank, they are equally likely to be chosen.
Thus, $\EM$ can be implemented efficiently
by observing that it chooses an output from the interval
$I_k = [x_k,x_{k+1})$ with probability proportional to
$|I_k| e^{-\frac{\eps}{2}|k-m|}$.
Conditional on $I_k$ being chosen in the first step, algorithm $\EM$
then returns a uniform random value in $I_k$.

\smallskip
\noindent
\textbf{Cell-based Method} is a heuristic proposed
in~\cite{Xiao:Xiong:Yuan:10}.
It imposes a fixed resolution grid over $C$ then computes the median
based on the noisy counts in the grid cells.
When applied to a hierarchical decomposition, a
fixed grid is computed over the entire data, then medians are computed
from the subset of grid cells in each node.
Cell counts have sensitivity 1. The accuracy of this
method depends on the coarseness of the grid relative to the data distribution.

\smallskip
\noindent
\textbf{Noisy Mean} is a heuristic from~\cite{Inan:2010:PRM}, which
replaces median with mean.
A private mean can be computed privately by computing a noisy sum (with
sensitivity $M$) and a noisy count (with sensitivity $1$), and
outputting their ratio.
If the count is reasonably large, this is a fair
approximation to the mean, though there is no guarantee that this is
close to the median.

\medskip
Provided the data is not too skewed, smooth sensitivity and
exponential mechanism have constant probability of choosing a good
split; i.e., the
noisy median has a constant fraction of the data points on each side.
We say that the data is not too skewed if it obeys the ``80/20 rule'':
the central 80\% portion of the data covers at least 20\% of the
entire data range $M$
(hence, it's not too concentrated).
Formally, $(x_{4n/5} - x_{n/5}) \geq M/5$.
For smooth sensitivity, we also require $n$ to be large enough that
$\xi n$ is at least a
small constant (where $\xi$ is as in Definition~\ref{smoothsens}).

\begin{lemma}
\label{smoothbound}
Let $C$ be such that $x_{4n/5} - x_{n/5} \geq M/5$.

(i) 
$ \Pr[ \SS(C) \in [x_{n/5}, x_{4n/5}] | n\xi \geq 4.03 ] > 0.5(1-e^{-\eps/4});$

(ii) $\Pr[\EM(C) \in [x_{n/5},x_{4n/5}]] \ge \frac16.$
\end{lemma}
\noindent
\textit{Proof.}
(i)
\begin{align*}
\textstyle
&\text{Observe that } \sigma_s & \leq \max (&\max_{0 \leq k < 2n/5} e^{-k\xi} (x_{m+k+1} - x_{m-k-1})
,
\max_{2n/5 \leq k \leq n} e^{-k\xi} (x_{m+k+1} - x_{m-k-1})
) \\
&& \leq \max( &(x_{4n/5} - x_{n/5}) , e^{-2\xi n/5} M)
\end{align*}
\noindent
Using our assumptions,
$
e^{-2\xi n/5} M < M/5 \leq (x_{4n/5} - x_{n/5}),$
so
$\sigma_s \le (x_{4n/5} - x_{n/5})$. Consider the case when the median is closer to $x_{n/5}$
than to $x_{4n/5}.$ Then the output of $\SS$ is within the desired range if
$0 \le (2\sigma_s/\epsilon) X \leq \sigma_s/2,$ i.e., $X\le \eps/4.$ Symmetrically, if the
median is closer to $x_{4n/5},$ the noisy median is in the desired range if
$0 \ge (2\sigma_s/\epsilon) X \geq -\sigma_s/2,$ i.e., $X\ge -\eps/4.$ Since $X$ is drawn from a
symmetric distribution, we conclude that
\[\textstyle\Pr[ \SS(C) \in [x_{n/5},x_{4n/5}] | \xi n \geq 4.03 ] > \Pr[ 0 \le X \le \eps/4]
 = \frac{1-e^{-\frac{\eps}{4}}}{2}.\]

(ii) Let $\alpha$ be the proportionality constant implicit in
Definition~\ref{expmech}. Let $E$ be the event that
$\EM(C) \in [x_{n/5},x_{4n/5}]$. Then
\[\begin{array}{rl}
\Pr[E]
  & \ge \alpha e^{-0.15\eps n} (x_{4n/5} - x_{n/5})
\\
\text{ and }
1-\Pr[E] = \Pr[ \neg E ] & \le \alpha e^{-0.15\eps n} M.
\end{array}\]
Using our assumption, we have $1 - \Pr[E] \le 5\Pr[E]$.  Hence $\Pr[E] \ge 1/6$.
\qed

\subsection{Balancing Privacy Budgets}\label{sec:budget}

As discussed before, 
the privacy guarantee of a data-dependent tree is obtained by
composing the individual privacy guarantees for medians and counts
along each root-to-leaf path, as in Lemma~\ref{epscomp}.

Let $h$ be the height of the tree and let $\A_i^m,$
$1\leq i\leq h,$ be the noisy median algorithms corresponding to the
internal nodes on a path, such that $\A_i^m$ is
$\eps_i^m$-differentially private.
Let $\A_i^c,$ $0\leq i\leq h,$ be the noisy count algorithms corresponding to the same
nodes, plus the leaf, such that $\A_i^c$ is
$\eps_i^c$-differentially private.
Then the PSD released
%
is $\eps$-differentially private, where
$$\eps = \sum_{i=1}^{h} \eps_i^m + \sum_{i=0}^h \eps_i^c$$
%
(assuming that sums are equal along each
path. Else, $\eps$ is the maximum sum along any path.)

\para{Median vs. count noise.}
In our experiments, we consider separate allocations of the
$\eps$ budget for the median computation, and for the count computation.
An important consideration is that larger values of
$\eps_i$ yield smaller noise (in a probabilistic sense). However, the
noise magnitude has different consequences on the overall accuracy of
the tree, for median vs. count.
Large count noise gives more uncertainty on the count of each
region in the tree which $Q$ touches.
By contrast, large median noise may
result in a skewed split in some internal node $u$.
Hence, the children of $u$ are unbalanced in
terms of number of points in their leaves,
as well as the size of regions they represent.
However, queries that
touch the descendants of $u$ may still perform well, provided the
respective noisy counts are accurate enough.
Still, we cannot neglect median finding entirely.
If the median does not divide the current point set so that there are
at least a constant fraction of points on each side, then we are
essentially wasting a level of the tree.
This becomes more of a challenge deeper in the tree, as the point set
sizes shrink.


Let $\eps_{\median}=\sum_{i=1}^{h} \eps_i^m$ and
$\eps_{\cnt}=\sum_{i=0}^h \eps_i^c$. We study different strategies for choosing these values such that
$\eps_{\median} + \eps_{\cnt}= \eps,$ by analyzing their impact on various query loads.

\para{Budgeting median noise.}
Once $\eps_{\median}$ is fixed,
we must distribute it among internal nodes. A simple strategy is uniform
budgeting, i.e., $\eps_i^m=\eps_{\median}/h.$
The hybrid tree approach, which switches to quadtree splitting after
$\ell$ levels implies setting $\eps_i^m = \eps_{\median}/\ell$ for $h-\ell < i \leq h$
and $\eps_i^m=0$ for $0 \leq i \le h - \ell$.


\para{Flattening the kd-tree.}
From Lemma~\ref{querybound} and the computation of query errors from Section~\ref{sec:counts},
it seems that a (2D) kd-tree built over the same
number of leaves as a quadtree has significantly worse accuracy.
The reason is that the kd-tree has twice the
height of the quadtree, so the noise budget is divided into twice as
many pieces.
To better compare quadtrees and kd-trees, we can ensure that both have
the same fanout of 4. We propose ``flattening'' the kd-tree: connect the root to its four
grandchildren, and recurse down the tree, skipping every other level. (This is akin to setting $\eps_i=0$
for every other level).
The bound on $n(Q)$ over a flattened kd-tree is now the same as for a quadtree
of the same height, following from Lemma~\ref{querybound}.
From now on, we assume all trees have fanout $f=4$.

\section{Further Enhancements}\label{sec:improve}

In this section we discuss how to apply two popular strategies---sampling and pruning---in the
context of differential privacy. Sampling can be used to significantly improve the running
time for computing data-dependent trees. Pruning can improve the query accuracy for both
data-independent and data-dependent trees.

\para{Sampling to reduce computation time.}
When data is large,
computing differentially private functions can be very time consuming,
even when the functions take near-linear time.
A natural technique is to compute the function on only a small sample
of the full data to speed up the computation.
Intuitively, sampling is very compatible with differential privacy:
the influence of any individual is reduced, since they may not be included in the sample.
The next result extends Kasiviswanathan {\em et
al.}~\cite{Kasiviswanathan:Lee:Nissim:Raskhodnikova:Smith:08}\footnote{Via the exposition at \url{http://adamdsmith.wordpress.com/2009/09/02/sample-secrecy/}}:

\begin{theorem}
Given an algorithm $\A$ which provides $\eps$-differential privacy,
and  $0<p<1$,
including each element of the input into a sample $S$ with probability
$p$ and outputting $\A(S)$ is
$2p e^{\eps}$-differentially private.
\end{theorem}

Treating $2e^{\eps}$ as a constant (it is between 2 and 5.5 for $0
< \eps < 1)$, it is sufficient to sample at a rate
of $\approx \eps'/10$ to achieve $\eps'$-differential privacy.
For large enough data, sampling at a rate of, say, $1\%$ and applying Laplace noise
with parameter $0.9$ achieves $0.1$-differential privacy---but it
processes a data set two orders of magnitude smaller.
We use this result to compute noisy medians for data-dependent
and hybrid trees, via the $\SS$ and $\EM$ methods from Section~\ref{subsec:medians}. We conclude that
sampling makes both methods an order of magnitude faster. It only marginally
deteriorates the accuracy of the exponential mechanism, and it
improves the accuracy of the smooth sensitivity-based noise. See Section \ref{exp:median}.

Sampling is not useful for noisy counts: computing a sample (independently per node) and computing the
exact count both require linear time. However, sampling introduces more inaccuracy in the noisy count. For the same
reason, the cell-based method for noisy medians (Section~\ref{subsec:medians}) is implemented without sampling.

\para{Pruning.}
Due to the uneven nature of the splitting process, it is possible that
both data dependent and data independent trees contain some nodes with
few or no points.
Dividing such sets can be counter-productive: the noise added to
the counts of their descendants may add up for queries which intersect
their region.
Instead, it is preferable to cut off the tree at this point.
This requires some care: the choice to stop cannot be based on the
``true'' count of the points, as this does not meet the differential
privacy requirement.
Instead, we can choose to stop if the noisy count estimated for
a node in the tree is less than threshold $m$.
In our setting, we choose to apply this pruning after the
postprocessing procedure of Section~\ref{sec:postprocess}, which
operates on a complete tree.
We observed that this improves empirically over retaining all nodes.


\newlength{\figwidth}
\setlength{\figwidth}{0.31\textwidth}

\section{Experimental Study}
\label{sec:experim}

We begin by studying design choices for each class of methods:
the choice of medians for data-dependent PSDs, different strategies
for budgeting and post-processing for data-independent PSDs.
We then study the effectiveness of query answering for the
different approaches.
Finally, we apply our findings to a particular application~\cite{Inan:2010:PRM}.

\begin{figure*}[t]
\centering
\subfigure[$\epsilon=0.1$]{\label{fig:quad-01}
\includegraphics[width=\figwidth]{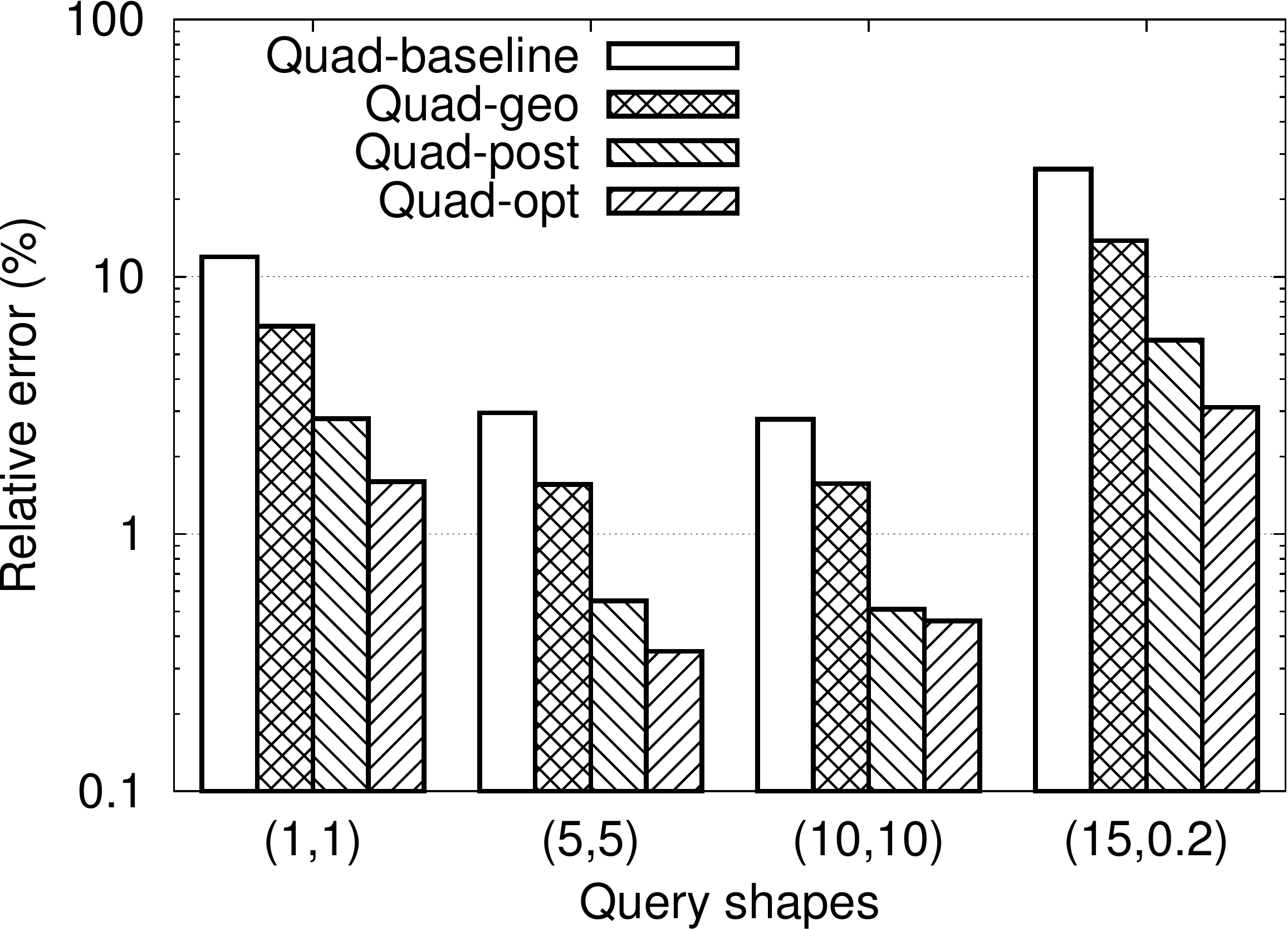}
}%
\subfigure[$\epsilon=0.5$]{\label{fig:quad-05}
\includegraphics[width=\figwidth]{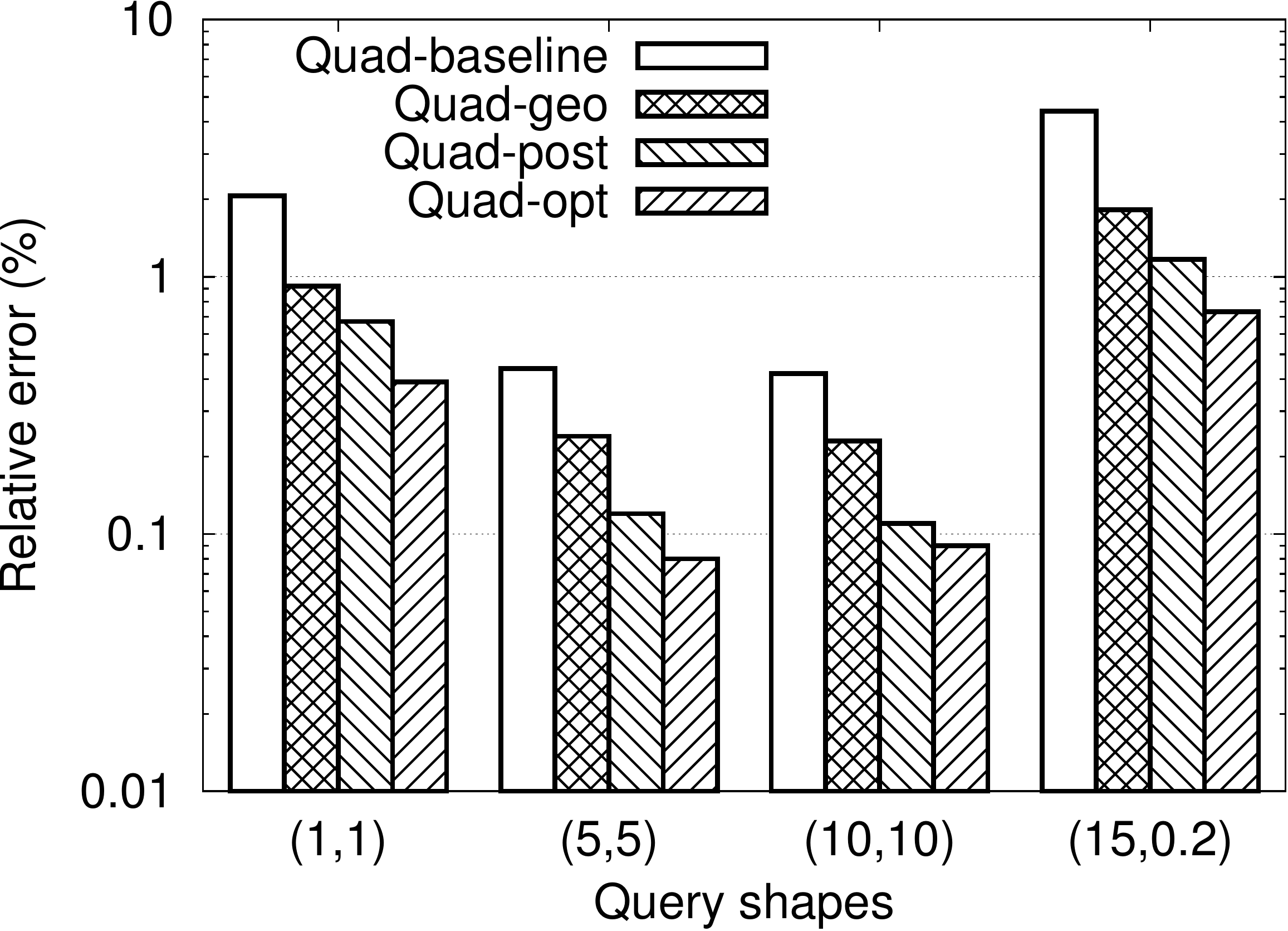}
}%
\subfigure[$\epsilon=1.0$]{\label{fig:quad-10}
\includegraphics[width=\figwidth]{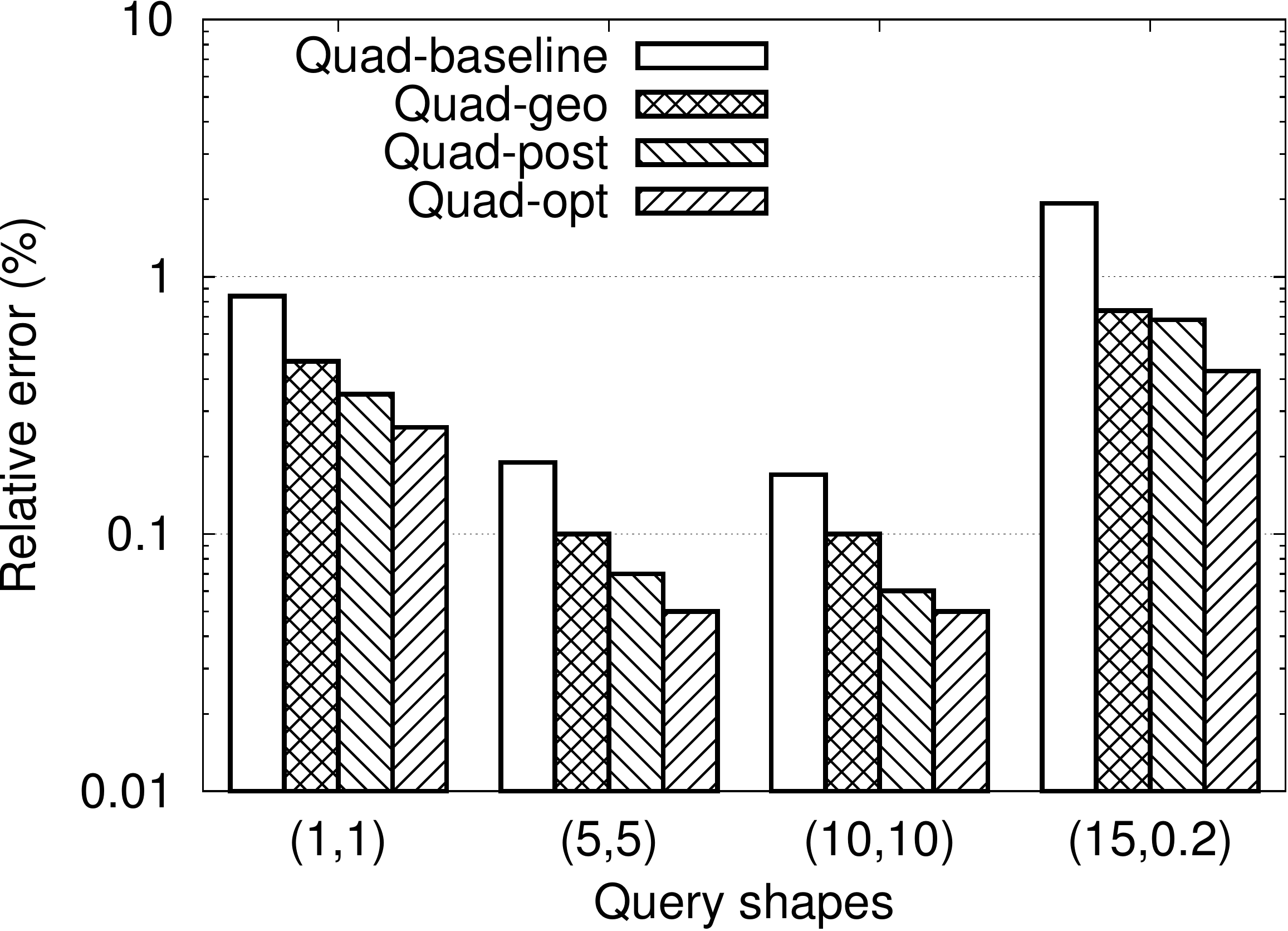}
}
\caption{Query accuracy of quadtree optimizations}
\label{fig:quadTrees}
\end{figure*}

\subsection{Experimental Environment}

To evaluate the accuracy of query answering of various PSDs,
we experimented on a mixture of real and synthetic data.
Real data was drawn from the
2006 TIGER/Line dataset\footnote{\url{http://www.census.gov/geo/www/tiger/}}.
Here, we present results using
GPS coordinates of road intersections in the states of Washington
and New Mexico.
This data represents a rather skewed distribution corresponding
roughly to human activity, so we treat it as the locations of
individuals which should be kept private.
The dataset has 1.63 million coordinates in the range
$[-124.82,-103.00]\times [31.33,49.00]$.
We conducted experiments on other data sets as well, including
synthetic 2D data with various distributions
and TIGER/Line data of different regions,
and obtained similar results.

We show results for rectangular queries
where query sizes are expressed in terms of the original data.
For example, since 1 degree is approximately 70 miles,
(15, 0.2) indicates a ``skinny'' query of 1050 $\times$ 14 miles.
We consider several query shapes;
for each shape we generate 600 queries that have a non-zero answer,
and record the median relative error.

All experiments were conducted on a 2.80GHz CPU with 8GB RAM, so the data
fits easily in memory.
We implemented our PSDs in Python 2.6 with scientific package Numpy.

\subsection{Experimental Results}

\para{Budget and post-processing.}
In Sections \ref{sec:counts} and \ref{sec:postprocess} we proposed two
techniques to improve query accuracy for PSDs:
geometric noise allocation, and the post-processing method.
We compare quadtrees with
geometric budgeting ({\em quad-geo}), post-processing ({\em quad-post})
and both combined ({\em quad-opt})
to the baseline approach of uniform budget with no post-processing
({\em quad-baseline}).
Figure \ref{fig:quadTrees} shows the relative errors
when all trees are grown to the same height, $h=10$.
Clearly, each optimization significantly improves the query
accuracy, and in combination they perform even better: the relative
error is reduced by up to an order of magnitude, especially when the
privacy budget is limited ($\epsilon=0.1$).
We observe the same improvement on other PSDs, so all subsequent
results are presented with both optimizations.

\para{Quality of private medians.}
\label{exp:median}
In Section \ref{subsec:medians} we described several approaches to private
medians: smooth sensitivity (SS),
exponential mechanism (EM), noisy mean (NM)
and the cell-based approach (cell).
To show their relative behavior, we compare them
on a synthetic one-dimensional dataset with $2^{20}$ data points
distributed uniformly within a domain of $[0,2^{26}]$
(the same
relative performance was seen on other distributions).
We build a binary tree structure with splits found by each mechanism,
and measure the average (normalized) rank error
of the private medians for each level.
In the worst case the noisy median may fall out of the data range
$[x_1, x_n]$ and have 100\% relative error.

\begin{figure}[t]
\centering
\subfigure[Private median quality]{\label{fig:noisy-median}
\includegraphics[width=0.33\columnwidth]{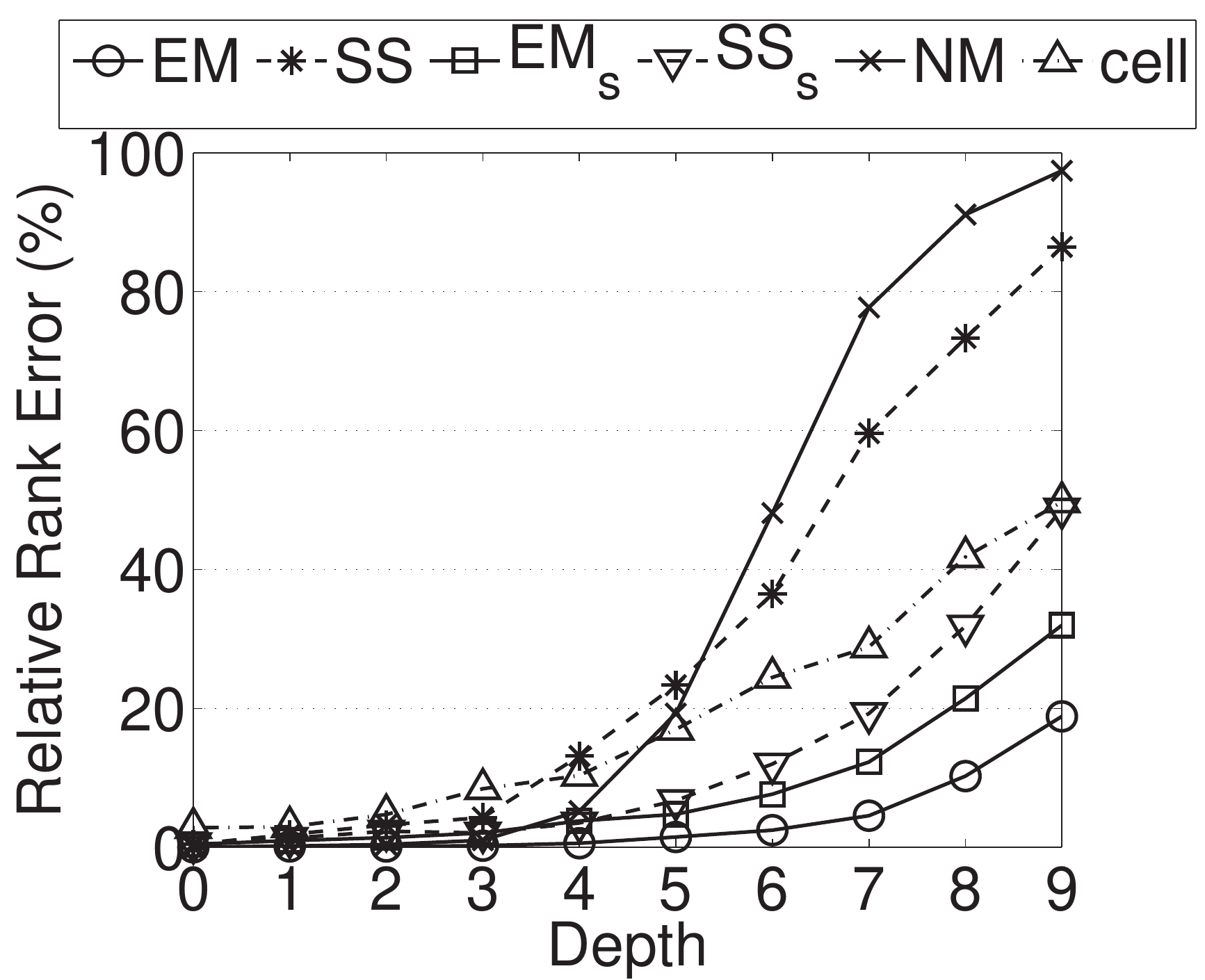}
}%
\subfigure[Time efficiency]{\label{fig:sample-time}
\includegraphics[width=0.33\columnwidth]{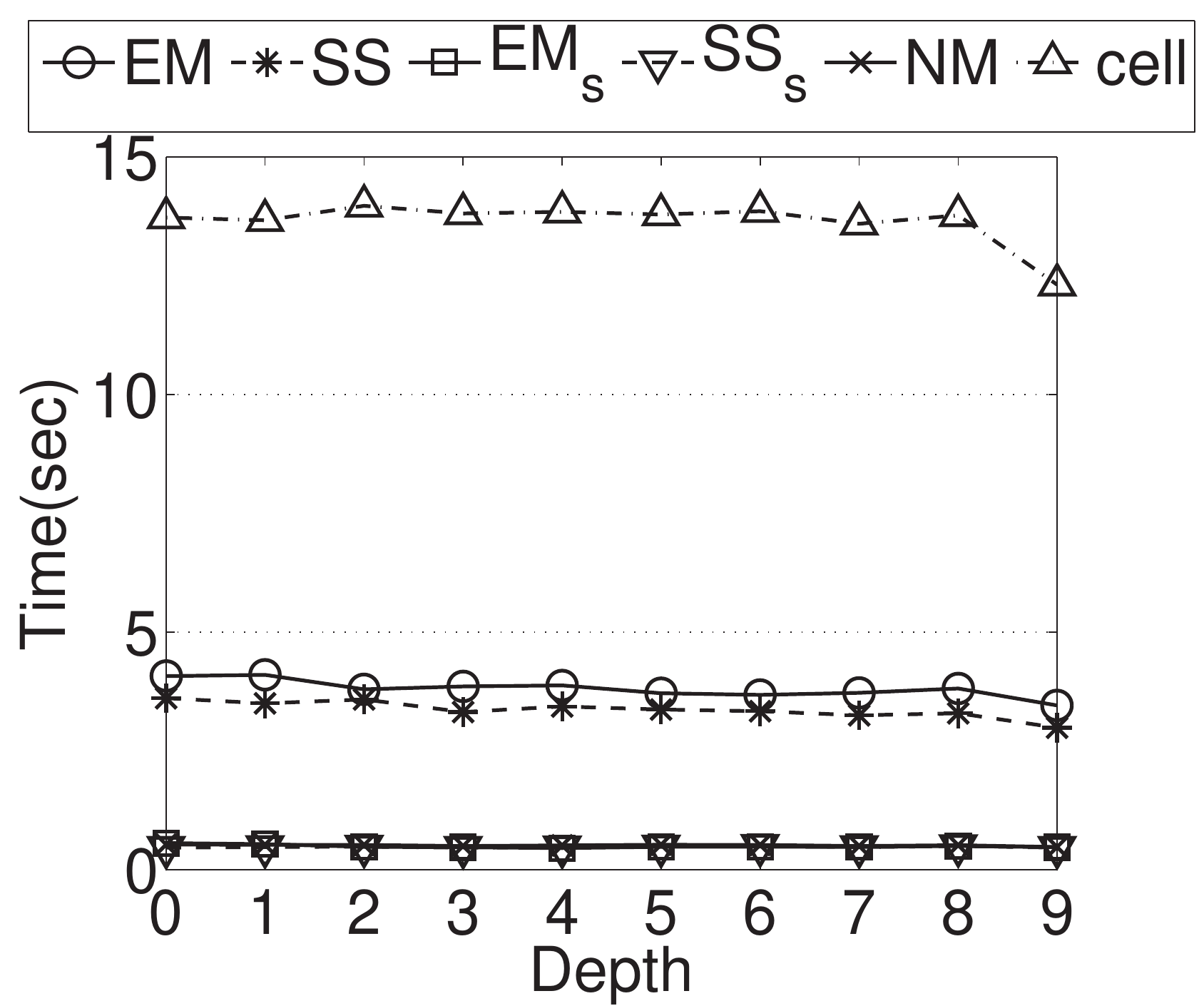}
}
\caption{Private medians accuracy and efficiency}
\label{fig:median}
\end{figure}

\begin{figure*}[t]
\centering
\subfigure[$\epsilon=0.1$]{\label{fig:kd-01}
\includegraphics[width=\figwidth]{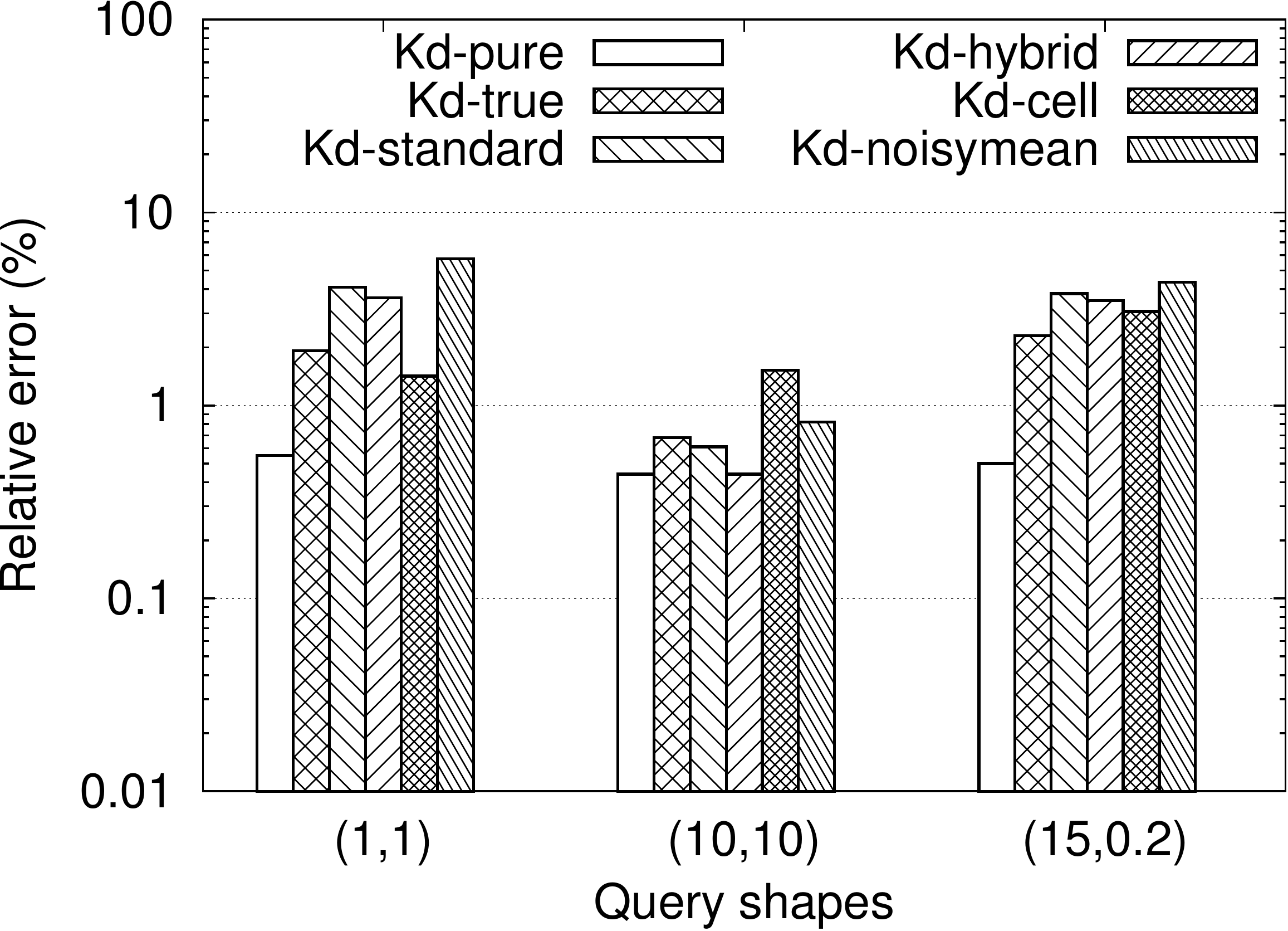}
}%
\subfigure[$\epsilon=0.5$]{\label{fig:kd-05}
\includegraphics[width=\figwidth]{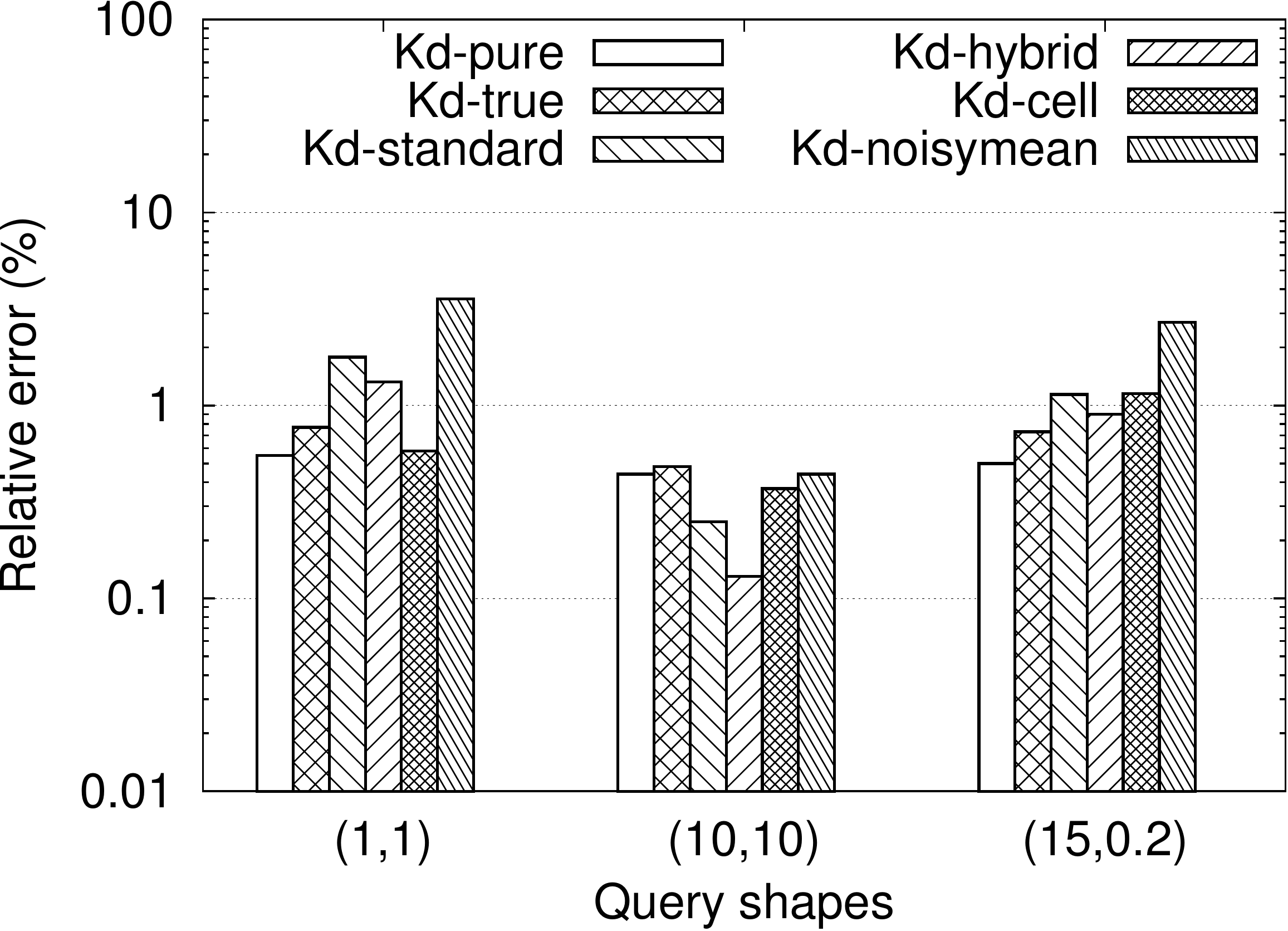}
}%
\subfigure[$\epsilon=1.0$]{\label{fig:kd-10}
\includegraphics[width=\figwidth]{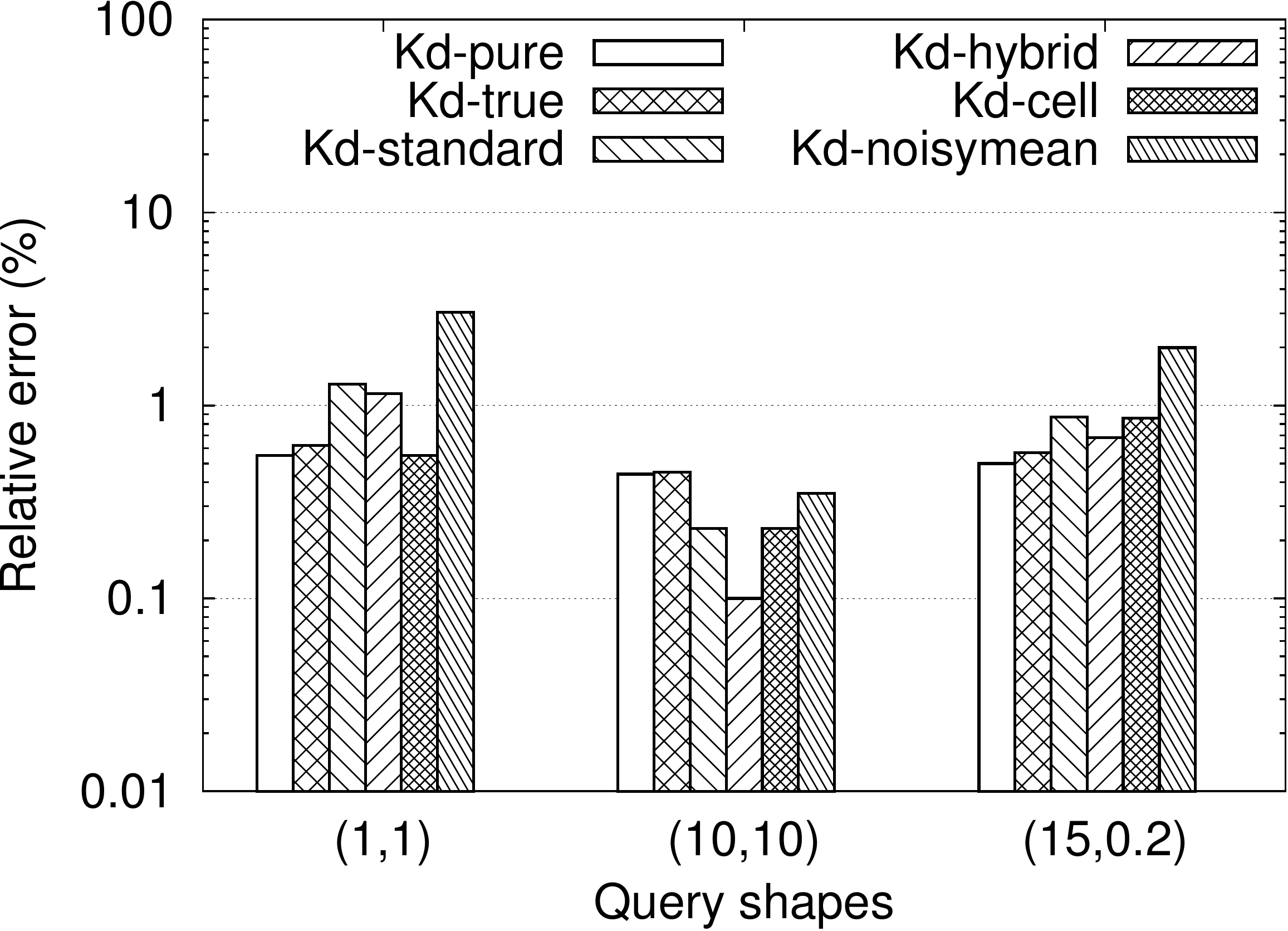}
}
\caption{Query accuracy for kd-trees}
\label{fig:kdTrees}
\end{figure*}

\begin{figure*}[t]
\centering
\subfigure[Query (1,1)]{\label{fig:height-10}
\includegraphics[width=\figwidth]{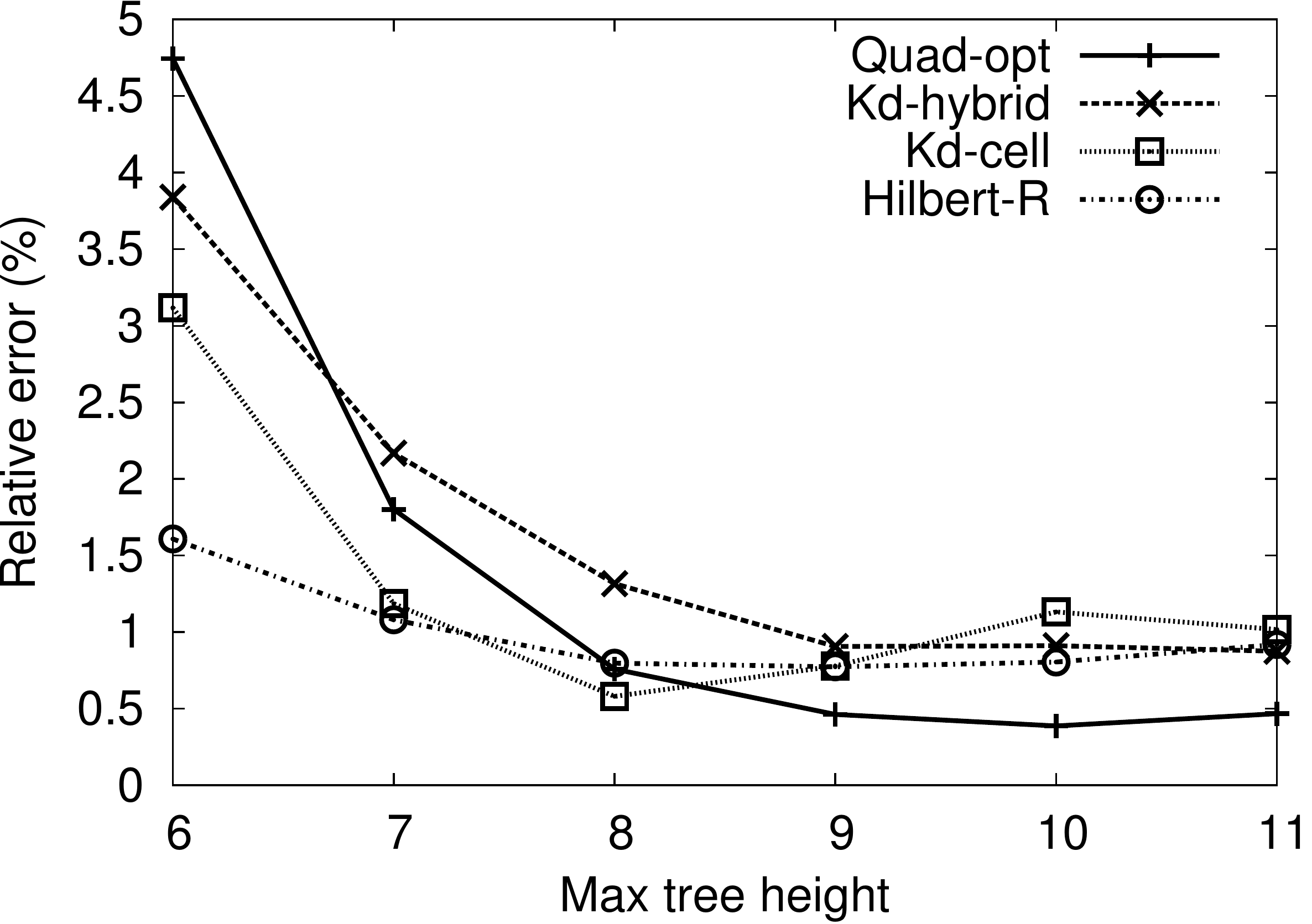}
}%
\subfigure[Query (10,10)]{\label{fig:height-100}
\includegraphics[width=\figwidth]{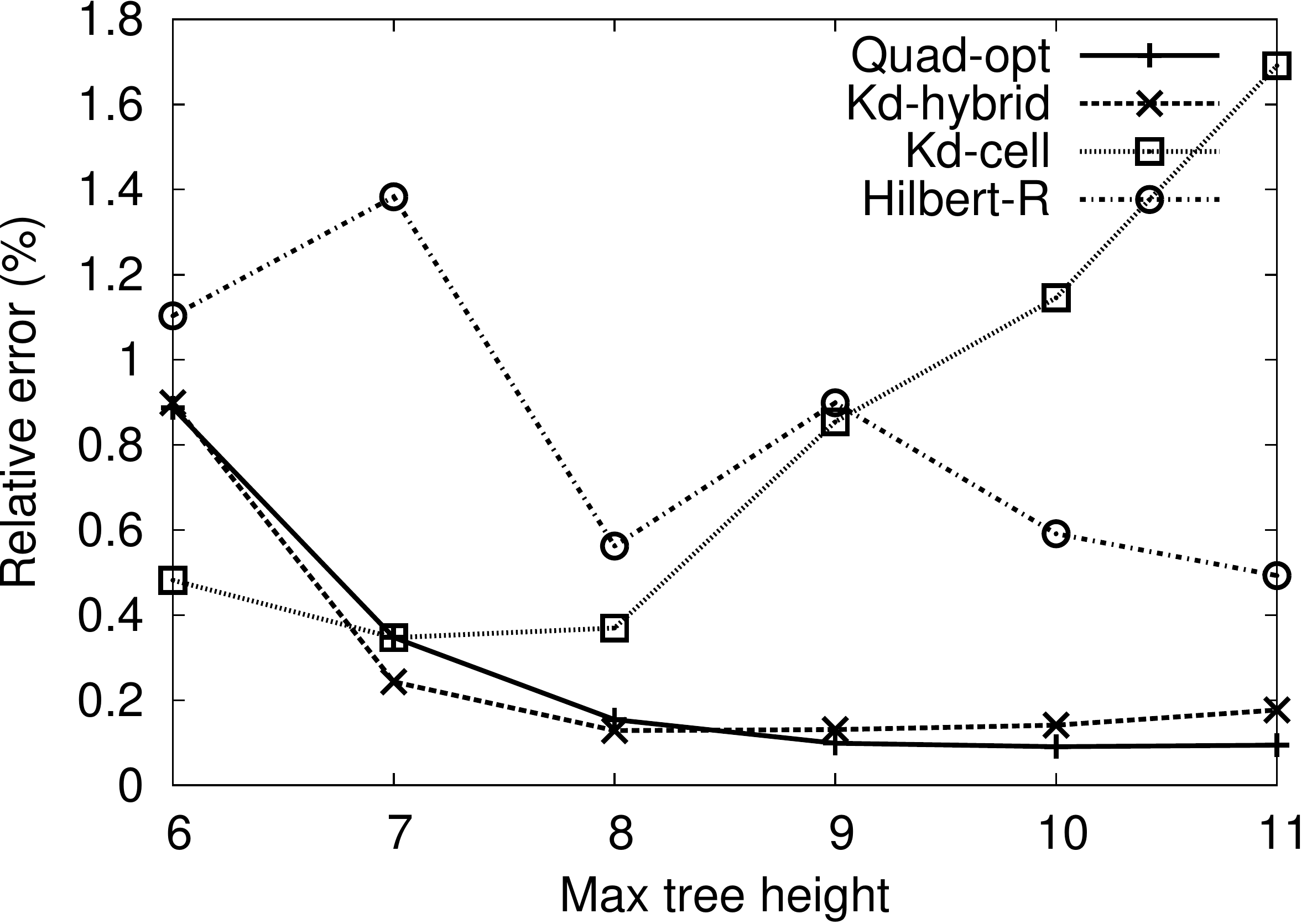}
}%
\subfigure[Query (15,0.2)]{\label{fig:height-150}
\includegraphics[width=\figwidth]{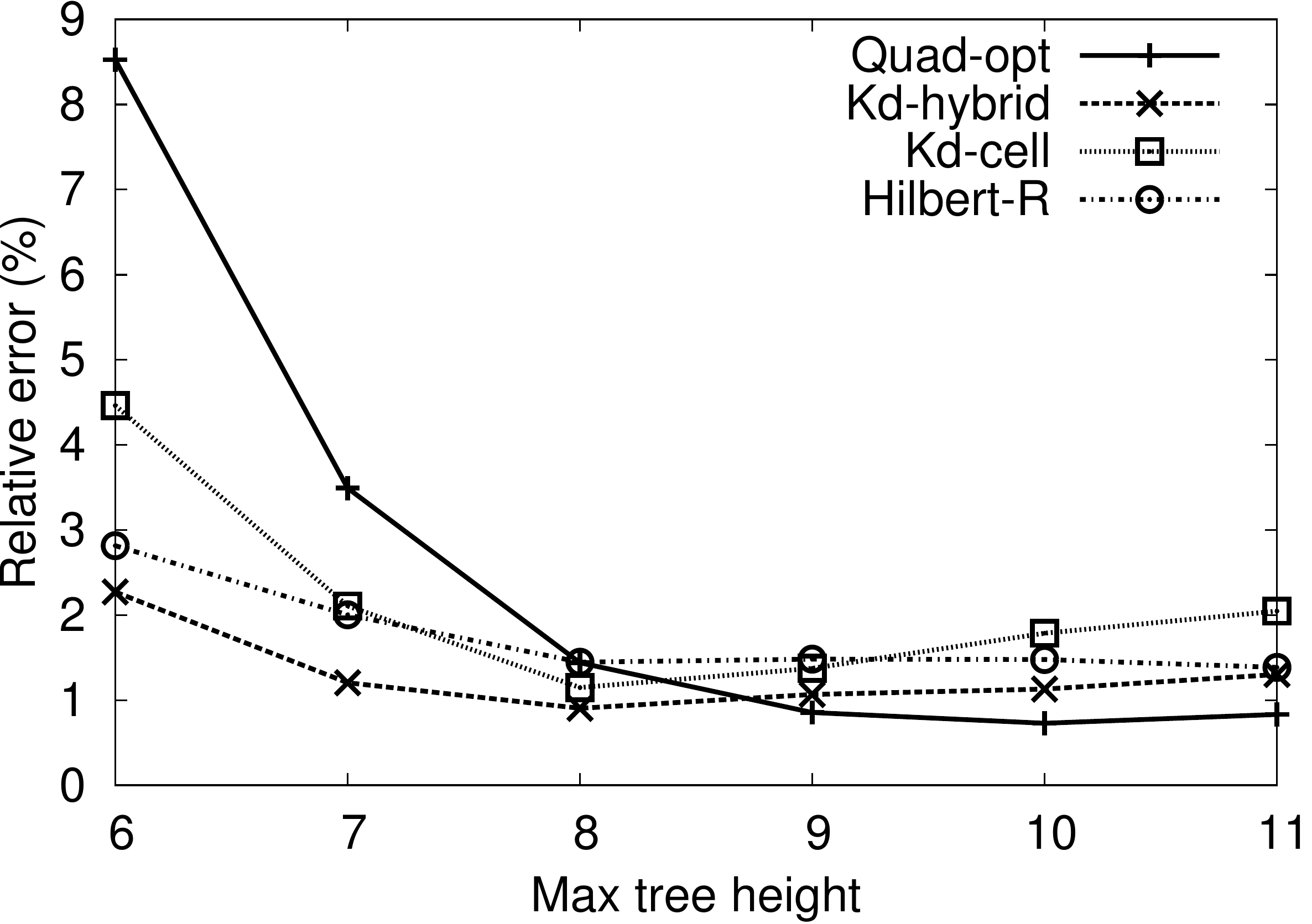}
}
\caption{Query accuracy comparison}
\label{fig:height}
\end{figure*}

Figure~\ref{fig:median} shows the accuracy and time for median
finding with privacy budget $\epsilon=0.01$ at each level (counting
from the root down). For SS, which is $(\eps,\delta)$-differentially private,
we set $\delta=10^{-4}$.
We also show the result of combining sampling with probability $p=1\%$
with SS and EM, obtaining methods SS$_{\text{s}}$ and EM$_{\text{s}}$
respectively.

We argued that efficiency could be improved by combining
either method with sampling.
The exponential mechanism (EM) is always the most accurate,
providing almost true medians for large data sizes
($2^{20}$ points at depth 0)
and medians with moderate error for smaller data sizes
(around $2^{11}$ points at depth 9) without sampling.
For smooth sensitivity,
the sampling (SS$_{\text{s}}$) approach
improves the accuracy, since the increase in privacy budget
outweighs the effect of fewer data points.
The reverse is true for EM$_{\text{s}}$: although the privacy budget
becomes about 50
times larger, the ranks are about 100 times smaller, so the
probability associated with elements far from the median is
correspondingly larger.
Nevertheless, as Figure~\ref{fig:sample-time} shows,
there is a commensurate performance improvement:
the sampling versions are about an order of magnitude
faster across all data sizes.
Although fast, the noisy mean approximation (NM) gives poor quality medians
for smaller data sizes.
The cell-based approach
(cell length $2^{10}$) has larger error for large data sizes
while being much slower than others.
Based on these results, we use the exponential mechanism as the default
method for noisy medians in all other experiments, and advocate the use of
sampling when the data under consideration is large.

\label{exp:query}

\para{Other parameter settings.}
We summarize our findings on parameter settings.
Detailed plots are omitted for brevity.

\smallskip
\noindent
{\em Hybrid trees.}
A hybrid tree switches from data dependent splitting  to
data independent splitting at a ``switch level'' $\ell$.
We found that switching about half-way down the
tree (height 3 or 4) gives the best result over this data set.

\smallskip
\noindent
{\em Median vs. count noise.}
As we increase  $\eps_{\median}$ and decrease $\eps_{\cnt}$,
we obtain more accurate divisions of the data, but less accurate
counts.
Although it may vary for different queries and PSDs,
in most cases the best results were seen when budget was
biased towards the node counts,
allocated roughly as $\eps_{\cnt} = 0.7\eps$ and $\eps_{\median} =
0.3\eps$, so we adopt this division.
\eat{
almost evenly
divided between $\eps_{\median}$ and $\eps_{\cnt}$.
For hybrid trees, allocating a small fraction of the budget still
gave good results. Going forward, we set the default budget
division to $\eps_{\cnt} = 0.4\eps$ and $\eps_{\median} = 0.6\eps$.
For hybrid trees, we use $\eps_{\cnt} = 0.8 \eps$ and $\eps_{\median} = 0.2\eps$.
}

\smallskip
\noindent
{\em Hilbert curve resolution}.
Recall that we can treat Hilbert R-trees as one-dimensional kd-trees, and so obtain PSDs.
First we set the {\em order} of the Hilbert curve.
Given the large domain and skewed data distribution,
a Hilbert curve of order 23 is
needed to differentiate each individual point in the dataset.
However, since the leaves of our PSDs should contain several points,
we found similar accuracy for all resolutions in the range 16 to 24,
and use curves of order 18 in our experiments.

\para{Comparison of kd-trees.}
We compare the accuracy of queries among
the kd-tree with EM medians ({\em kd-standard}),
the hybrid kd-tree ({\em kd-hybrid}),
the cell-based kd-tree proposed in \cite{Xiao:Xiong:Yuan:10}
({\em kd-cell}) with cell length 0.01
and the noisy mean based kd-tree from \cite{Inan:2010:PRM} ({\em
kd-noisymean}).
To see ``the cost of privacy'',
we also include some results on non-private structures:
the exact kd-tree ({\em kd-pure}) and the
kd-tree which uses exact medians but noisy counts ({\em kd-true}).
In all experiments, the kd-trees have the same (pre-determined)
maximum number of levels $h=8$ (with fanout 4).
We also use a pruning condition based on frequency: we
remove descendants of any node whose noisy count is below a threshold
$m = 32$.

Figure~\ref{fig:kdTrees} shows the result of comparing kd-tree variants
across a range of privacy budgets ($\eps = 0.1, 0.5, 1.0$).
We first observe that even a pure kd-tree without noise of any
kind introduces some error because of the uniformity assumption for leaves
that intersect, but are not contained in the query.
Comparing \textit{kd-true} and \textit{kd-pure}, we see that
adding noise to counts does not dramatically degrade query
accuracy: it remains below 1\% relative error.
Rather, it is the noise in the choice of medians which seems to have
the greater impact on query accuracy, as is evident from the results
for the PSDs.
Since the splitting points do not evenly divide the data, particularly
at greater depths in the tree, there are some leaves with a much
larger number of points than in the exact trees.
This explains the weak performance of {\em kd-noisymean}, whose choice of
splitting points was seen to be poor in Figure \ref{fig:noisy-median}.

The {\em kd-cell} approach has good performance on queries which are
small and square.
However, as the query area grows, the method looses its edge, and is
dominated by {\em kd-hybrid}.
Overall, the accuracy of query answering from all cases
is good, with small relative errors of less than 10\% even
for high privacy ($\eps=0.1$).
Of the kd-tree variations, the kd-hybrid seems the most reliably
accurate.

\para{Comparison of PSDs.}
Figure \ref{fig:height} shows the query accuracy for the
best performing instances of the representative methods:
optimized quadtree, cell-based kd-tree, hybrid kd-tree and Hilbert R-tree
on three queries of different shapes.
In this set of experiments we varied the depth of the trees
($h=$6 to 11) while keeping the privacy budget fixed at $\epsilon=0.5$.
At depth 10 the optimized quadtree is able to provide the best accuracy
of all.
It seems that the ability to devote all of the privacy budget to noisy
counts for quadtrees outweighs the ability of kd-trees to
more evenly split the data via private medians.
However, the hybrid kd-tree at height $h=8$ obtains about the same
accuracy as the quadtree at height $h=10$  for the larger queries.
The cell-based kd-tree, which first imposes a uniform grid over the domain,
provides good accuracy on small square queries since the node shape
matches the query shape.
However, {\em kd-cell} performs worst
on the larger queries, which require probing more nodes.
We also note that the private Hilbert R-tree has
comparably good performance on some queries, but much higher errors on
others.


\begin{figure}[t]
\centering
\subfigure[Construction time]{
\includegraphics[width=0.33\columnwidth]{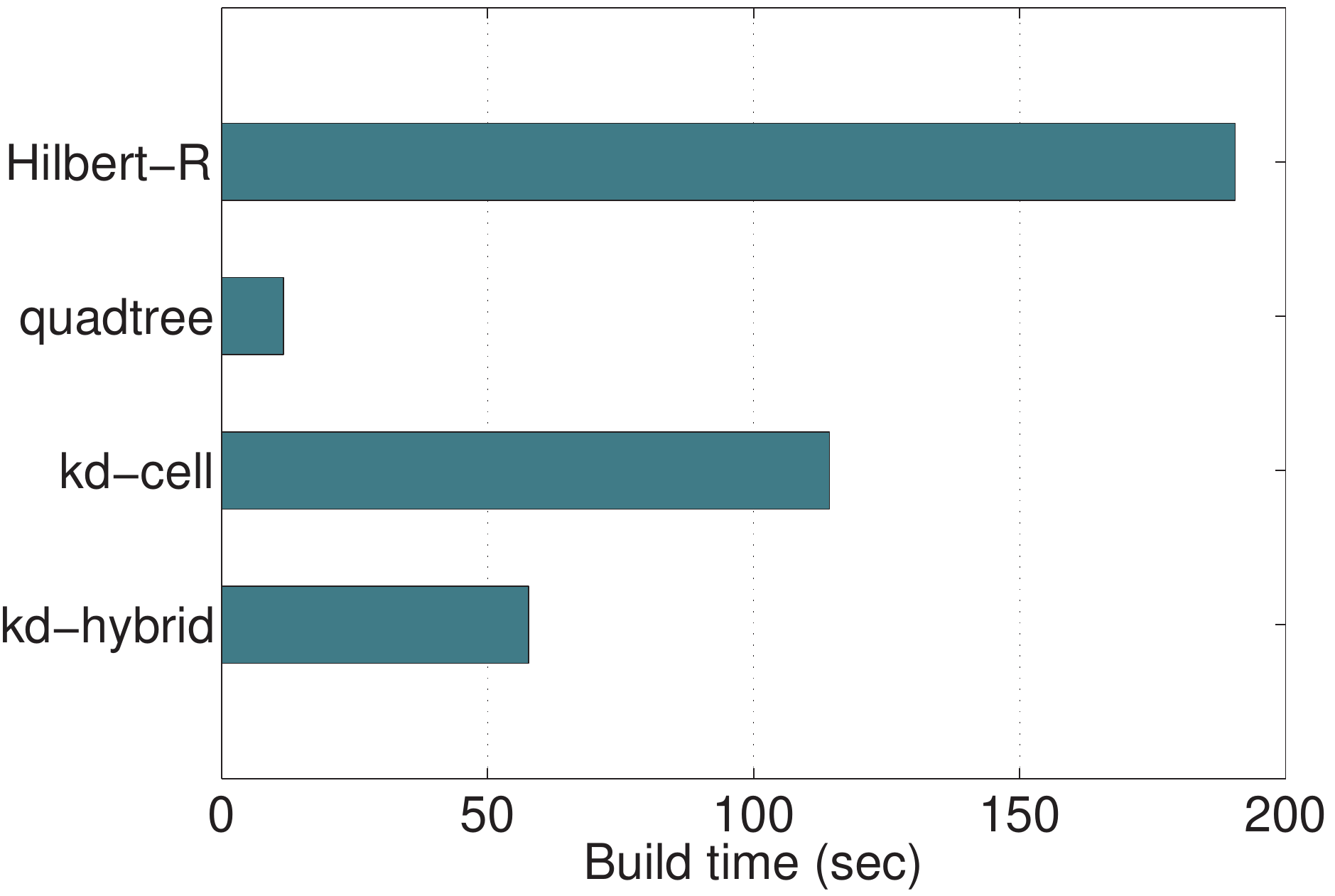}
\label{fig:time}
}%
\subfigure[Results for private record matching]{
\includegraphics[width=0.33\columnwidth]{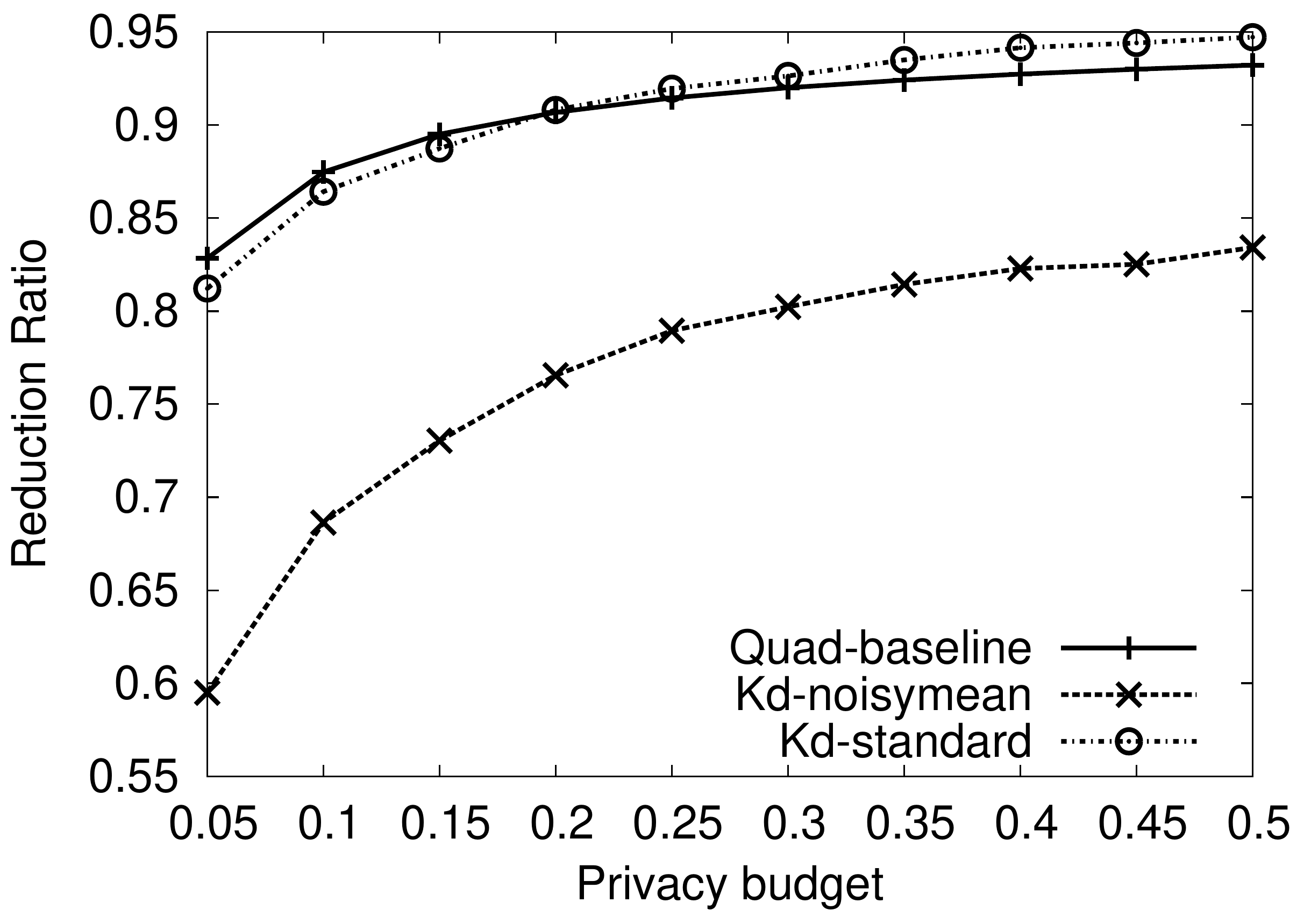}
\label{fig:edbt}
}%
\caption{Time efficiency and results on private record matching}
\end{figure}

\para{Scalability.}
Figure \ref{fig:time} illustrates the time taken to build each spatial
decomposition using our prototype code.
For the data sets considered, PSD construction time was not a
bottleneck, since we typically consider this to be a one-time cost.
In general,
the structures which only divide the domain take much less time than the data
dependent approach, while hybrid kd-tree lies in-between, at around a minute.
The cell-based kd-tree also takes a long time, due to first
materializing noisy counts of many cells, then building a tree on top
of them.
The Hilbert R-tree takes even more time,
due to the higher cost of encoding and decoding coordinates in Hilbert
space.
The count post-processing step adds only a few seconds to the overall
cost.

\subsection{Application to Private Record Matching.}
\label{exp:edbt}

The {\em kd-noisymean} tree was originally introduced to solve a
 private record matching problem---see the full details
in \cite{Inan:2010:PRM}.
Here, a differentially private data structure is used to focus in on
areas of potential overlap between datasets before an expensive secure
 multiparty computation (SMC) procedure is invoked.
The metric of interest is the {\em reduction ratio}, which measures
how much SMC work is saved relative to the baseline of no elimination, so
bigger is better.
We obtained the code from the authors of \cite{Inan:2010:PRM}
and ran the same experiments on the same data.
Figure~\ref{fig:edbt} shows reduction ratio
for {\em quad-baseline}, {\em kd-noisymean} and {\em kd-standard} (in this
application, all count budget is allocated to leaves and thus
post-processing does not apply).
We observe that the new kd-tree approach we have proposed can
improve appreciably over the two methods tried in~\cite{Inan:2010:PRM}.
Note that improving reduction ratio from 0.93 to 0.95 represents 28\%
less SMC work.

\section{Concluding Remarks}

We have presented a comprehensive framework for differentially private spatial
decompositions, and shown how to produce private versions of many
popular methods, such as quadtrees, kd-trees and Hilbert R-trees.
We have proposed novel techniques for setting hierarchical noise parameters
in a non-uniform manner that minimizes query error. Further, we have developed
a post-processing technique that re-computes node counts based on the
initial noisy counts to optimize query accuracy. This new technique
applies to a large class of other settings in privacy.

In the process of computing private data-dependent decompositions,
we have provided the first survey of techniques for computing
private medians and derived theoretical results that give insight into
the expected behavior of the two most accurate methods. We
have shown
how to combine all these results (plus other techniques such
as sampling) into a single whole that achieves the desired privacy guarantee.
In our ongoing work, we study other settings: sparse categorical
data~\cite{Cormode:Procopiuc:Srivastava:Tran:12}, and higher dimensional data.

Our experimental study is consistent with our theoretical insights,
and shows that each of our techniques leads to significant improvements
in query accuracy, running time or both. For most PSDs,
we achieve relative query errors in single-digit percentages, while
guaranteeing strong differential privacy.
We conclude that PSDs
represent an efficient and accurate way to release spatial data privately.


\para{Acknowledgments.}
We thank Adam Smith and Michael Hay for some helpful discussions.
Shen and Yu are partially supported by the National Science Foundation under the award CNS--0747247 and the NCB-Prepared project.

\bibliographystyle{abbrv}
\bibliography{spatial}

\end{document}